\documentclass[a4paper,12pt]{article}
\usepackage{color}
\usepackage{multirow}
\usepackage{amsmath}
\usepackage{xcolor}
\usepackage{graphicx}
\usepackage{hyperref}
\usepackage{url}
\usepackage[latin1]{inputenc}

\begin{document}
\date\today
\newcommand{\met}{\ensuremath{{\not\! E}_{T}}}
\newcommand{\pt}{p_{T}}
\newcommand{\bul}{\textbullet}
\newcommand{\be}{\begin{equation}}
\newcommand{\ee}{\end{equation}}
\newcommand{\br}{\begin{eqnarray}}
\newcommand{\er}{\end{eqnarray}}
\newcommand{\lumi}{{\cal L}}
\newcommand{\invfb}{fb^{-1}}

\begin{titlepage}
\begin{center}
{\LARGE\bf
Jet Cross-Section Measurements In CMS \\[5mm]}
\bigskip
{\large\sf Sanmay Ganguly\footnote[1]{sanmay@tifr.res.in}  
and
 Monoranjan Guchait\footnote[2]{guchait@.tifr.res.in} } \\
\vspace{1cm}
\date{\today}
Department of High Energy Physics, \\ 
Tata Institute of Fundamental Research,  \\
\hspace*{0.1in} 1, Homi Bhabha Road, Mumbai 400 005, India. 
\end{center} 
\hspace{1cm}
\begin{abstract}

The Large Hadron Collider (LHC) experiment has successfully completed data taking at
center of mass (COM) energy 7 TeV in 2011 and very recently for 8 TeV. Measurement of cross sections
predicted by the standard model were the main tasks in the beginning.
The inclusive jet cross section and dijet mass measurement is already
done at 7 TeV energy by Compact Muon Solenoid (CMS) detector with integrated luminosity
5 fb$^{-1}$.
In these measurement one needs to understand and measure precisely
the kinematic properties of jets which involve many theoretical and
experimental issues. The goal of this article is to discuss all these
issues including jet measurements in CMS and subsequently review the inclusive
jet
cross section and dijet mass measurement in CMS at 7 TeV with
integrated luminosity 5 fb$^{-1}$. The measurements, after unfolding the data, are also compared 
with the next leading order (NLO) theory predictions,
corrected for the non-perturbative (NP) effects,
for five different sets of parton distribution functions (PDF). It is observed
that the measurements, for both cases, agree with the theory prediction
within
$\sim$8-10\% depending on transverse momentum ($p_T$) and dijet invariant mass ($M_{jj}$) of jets. 

\end{abstract}
\hspace{1cm}

Pacs Numbers: 12.20.-m, 12.20.Fv, 11.15.Bt \newline

Keywords: QCD, Jet Algorithm, Cross Section, CMS
\end{titlepage}

\tableofcontents
\section{Introduction}
Quantum chromodynamics (QCD) describes the theory of strong interaction
among colored particles, viz. quarks and 
gluons ~\cite{qcd1,qcd2,qcd3}. It is one of the very well understood 
theory of subnuclear physics which has been tested in various
experiments with a high accuracy~\cite{qcdtest}. 
Since QCD deals with quarks and gluons, hence, in any hadron collider 
machine all partonic 
interactions are dominantly governed by it, in particular perturbative 
QCD (pQCD) plays an important
role in describing the parton dynamics.
In the hard scattering process, the partons, immediately after production,
fragment and hadronizes forming a cluster of collimated energetic colorless 
particles, hadrons. A clustering algorithm is applied on these particles to 
form a collection of particles which are called jets, the experimental 
analogue of partons
and one of the key observable in the theory of QCD.
Although jets are formed out of the fragmentation of colored partons, 
nevertheless it is colorless
and a very robust observable in QCD. 
Naturally, any measurement of jet energy and momenta 
will lead to a close estimation of the dynamical properties 
associated with the partons. 

Jet observables carry kinematic informations of interactions taking place 
at the parton level. 
For example, in the inclusive jet production,
where jets are produced by parton-parton interaction, the momenta of jet 
and the corresponding
parton momenta are
almost identical in the partonic center of mass (COM) frame. 
In this case a study of inclusive jet production  gives an estimation of 
distribution of
partons within the proton. Moreover, since inclusive jet production is mainly
controlled by QCD, therefore, measurements of jets and related many
observables, like 
event-shapes~\cite{Banfi:2010zza,Aad:2012np,Khachatryan:2011dx},
jet shapes~\cite{Ellis:2010rwa,Chatrchyan:2012mec,PhysRevD.83.052003} 
are employed to test various features of QCD. Inclusive jet cross section 
is also one of 
the important measurement which enables to  
measure the value of strong coupling constant ($\alpha_{S}$) and its 
running with
energy~\cite{PhysRevLett.88.042001,Chatrchyan:2013txa,alphas:atlas}. 
In addition, jets are 
also produced from heavy standard model (SM) particles like 
W/Z bosons or top quarks decay to quarks accompanied with other objects 
like leptons 
and photons. Therefore, an accurate reconstruction of jets are 
required to reconstruct the mass
of the parent particles. For instance, a precise 
estimation of top quark mass in its full hadronic decay depends how accurately 
jets are
 reconstructed
~\cite{PhysRevLett.79.1197,top:cms,top:atlas}.
 Moreover, many 
beyond standard model (BSM) particles predict some hadronic 
resonances for which precise measurements of jets are very crucial
~\cite{jetsreso,top:atlas1,PhysRevD.87.072002,Harris:2011bh}. 
One of the very popular BSM candidate, the supersymmetry predicts 
signal accompanied with a lot of jets along with other 
objects.
The SM processes with identical final states consisting lot of jets are
the dominant backgrounds corresponding to various BSM signals. The searches 
for BSM require 
a very good understanding and measurements of kinematic properties of
jets which are used to isolate signal from background events~\cite{Chatrchyan:2013lya,susy_atlas}. 
BSM particles are anticipated
to be heavier than the standard model particles. The decay product of 
these particles will be
boosted and the decay products will be confined within a narrow cone. To resolve these particle
kinematics, jet substructure techniques has also been used for sometime.
~\cite{PhysRevLett.100.242001,atlas_substructure}. \newline

In general, to analyze high energy phenomena,
we use event generators which emulate real experiments
starting from matrix element based calculation. These event generators are based on different 
QCD Monte Carlo (MC) models. As mentioned before, partons produced due to hard collisions, fragment
and hadrnoize leading to a showering phenomena which occurs at very low energy scale much below the pQCD
regime where $\alpha_{S}$ becomes too large. The role of event
generators are to implement the model of showering of particles following various methods.
The energy scale at which showering takes place, $\alpha_{S}$ becomes larger than unity and
hence this process is non-perturbative (NP) in nature. 
As a consequence, the characteristics of final state particles, in particular jet formations
are to certain extent influenced by this NP models. Therefore, any  observables based on jets, like event 
shapes~\cite{Banfi:2010zza,Aad:2012np,Khachatryan:2011dx}
measurement are ready to use to constrain this NP models. In summary, starting from precision study of 
strong interaction dynamics to background estimation to isolate 
BSM signals, jet study plays one of the 
most pivotal role in high energy physics experiment. \newline

Reconstruction of jets is a challenging issue from both experimental
and theoretical standpoint. The clustering or grouping of hadrons originating 
from partons are performed by following certain techniques which are called jet 
algorithm~\cite{Tkachov:1999py}. Theoretically, the formation of jets 
are suffered by infra-red (IR), both soft and collinear, and ultra-violet (UV) divergences. The 
construction of jet algorithm depends on how the issues
of divergences are resolved. In order to deal with these non-trivial issues, various methods are
proposed leading to different types of jet algorithms. 
In jet reconstruction by a given jet 
algorithm, one of the main input parameter is the value of jet radius R
defined in the azimuthal and pseudo-rapidity 
plane~\cite{Sterman:2004pd,Salam:2008qq}. The choice of the value of R decides the amount of hard
scattered partons clustered into jets. \newline

In this article we discuss the inclusive jet cross section and dijet 
invariant mass measurement with the  
Compact Muon Solenoid (CMS) detector at 7 TeV LHC run in 2011 with integrated
luminosity 5fb$^{-1}$. A comparison with the various theory predictions 
are also discussed in detail.  
At this high energy it is possible to reach to 
comparatively more lower region of Bjorken ($x\sim 10^{-3}$) and higher value
of $Q^2$. These measurements enable us to test various implications of QCD at 
this new regime of the phase space. It is worth to mention here that these 
type of studies are also performed in earlier experiments, like in Tevatron 
by D0~\cite{incld0,dijetd0}, CDF~\cite{inclcdf} and  much before 
by UA2~\cite{inclua2},
HERA~\cite{inclhera,inclheraktakt} collaborations.  
Very recently the ATLAS group in LHC experiment reported their 
measurement with 7 TeV datasets~\cite{inclatlas}. Here we present results
based on the recent measurements by CMS for 7 TeV energy which are reported 
in Ref.~\cite{inclcms,dijetcms,CMS-PAS-QCD-11-004}  \newline

We organize this article as follows.
In Section 2
we briefly discuss about theoretical issues related with
jet definitions and jet reconstruction algorithms followed
by a short description of detector and jet reconstruction techniques in
CMS experiment in Section 3.
We discuss the event and jet selection requirements
and the variables which are measured in Section 4. The estimation
of different uncertainties on the measured spectrum are discussed
in Section 6 after discussing the unfolding procedure in Section 5.
A detailed study of comparison between the measurement
and next to leading order (NLO) theory prediction is presented in Section
7.
Finally after discussing results in Section 8 we summarize in Section 9.

\section{Jet Algorithm}
In QCD studies jets are basically the transformed states of partons
to hadrons through the hadronization process i.e. a collection of spray
of particles exactly what happens in cosmic ray events.
Obviously, in any QCD measurement where partons are produced the first 
task is to reconstruct jets out of hadrons or calorimeter towers. 
Here the reconstruction of jets means 
the clustering of stable particles or
the calorimeter cells, in which energy is deposited by the produced 
stable particles,
following certain rules, which is called jet algorithm. 
More precisely, jet algorithm defines the strategy by which clustering 
can be performed. In formulating these prescriptions QCD plays a dominant 
role. Historically, the jet cross section was calculated dates back
to late 1970's after the discovery of asymptotic 
freedom~\cite{Gross:1973ju,Politzer:1974fr} which is
the main essence of QCD. The jet level cross section from partonic level was
calculated in the electron-positron annihilation. In this context first time
jet reconstruction was discussed by several 
authors
before the discovery of gluon~\cite{Hanson:1975fe,Ellis:1976uc,PhysRevLett.39.1436,DeRujula:1978yh}.
With this jet algorithm, another prescription is required to 
obtain the kinematic properties of jets out of this clustered objects,
which is the recombination scheme. It prescribes how to 
recombine jet constituents i.e. energies of calorimetric cells or momentum of
particles to build up finally the jet momentum and energies.
Therefore, any jet algorithm requires certain input parameters and a
recombination schemes, which together is called jet definition.
One of the very important issue requires to be addressed while constructing
a jet definition is to make the algorithms free from any kind of
divergences originating from collinear or soft branching of partons.
Here we note that, as mentioned before, one of the striking feature occurs
in the prediction of any high energy physics processes  based on 
fixed order pQCD calculation is the infra-red and collinear (IRC) divergence.
The IRC appears 
because of emission of two partons at a very small angle
with respect to each other (collinear) or momenta of one of the emitted
parton is very small (infra-red). It leads to divergent matrix elements
at the tree and as well as loop level making it IRC
unsafe. In reality, a algorithms should be free from the effect of the
IRC singularity. Therefore, from all these considerations an ideal jet
definition should offer \cite{Huth:217490,runiijet}: \newline
\\
\bul  It should be simple to implement in an experimental analysis.
\\
\bul  It should be easy to implement in theoretical calculations.
\\
\bul  It should be defined at any order of perturbation theory.
\\
\bul  It should yield finite cross section at any order of perturbation
theory. 
\\
\bul  It should be insensitive to infrared singularities.\\ 
\\
Historically, the first jet algorithm was developed based on cone
algorithm\cite{PhysRevLett.39.1436}. This cone algorithm prescribes to form 
jets by clustering the set of particles whose trajectories lie 
within a radius R in $\eta \times \phi$ space, where $\eta$ is the 
pseudo-rapidity defined to be, $\eta$ = -ln$(\tan\theta/2)$ and $\phi$ is the azimuthal 
angle in the x-y plane.{\footnote {In collider, beam direction is 
assumed to be in z direction and x-y plane is perpendicular to the beam.
$\theta$ is the angle between the particle momentum direction and +z direction.}}. 
The cone algorithm, more precisely iterative cone (IC) algorithm first 
decides a seed particle and the corresponding direction as a seed direction, 
then combine the momenta of all other particles if they are within a radius,\\
\begin{equation}
\Delta R= \sqrt{\Delta\phi^{2} + \Delta \eta^{2}}
\label{eq:dR}
\end{equation}
where $\Delta\eta $ and $\Delta\phi$ are the differences of pseudo-rapidities and
azimuthal angles respectively between the corresponding pair of particles. Then the new direction is
regarded as seed direction and iterate this calculation till the stable
direction is reached. This algorithm are used in jet reconstruction
in early days by UA1~\cite{ua1} experiment and also in early part of Tevatron
experiment~\cite{PhysRevD.78.052006}. However this IC algorithm is suffered by many
problems, and one of the major one is that it is IRC unsafe i.e due to the soft 
emission or collinear splitting, the properties of hard jets in the event 
changed. In order to cure this problem, 
few variations of IC method were introduced, namely,
midpoint cone~\cite{conealgo} and SisCone~\cite{conealgo}. 
However the detail discussion of pros and cons of various jet algorithms, which is out of scope of the 
present article can be found 
in the literature~\cite{Sterman:2004pd,runiijet}.
It is also to be noted that in the formalism of jet algorithm, along with IRC safety issue in 
jet formation, another non trivial issue is the computational power. 
Therefore, always attempts were there
to develop more and more elegant techniques addressing all requirements to perform jet reconstruction. 
In this endeavor, quite a few interesting algorithms were developed, 
namely JADE~\cite{jade} algorithm, $k_T$ algorithm~\cite{Catani:1992zp},
anti-$k_T$~\cite{antikt} algorithm and 
Cambridge-Achen(C/A)~\cite{caalgo} algorithm. \newline

Currently, in hadron collider experiments, like Tevatron and LHC,
the $k_T$~\cite{Catani:1992zp,Catani1993187} and anti-$k_T$~\cite{antikt} algorithms has become 
very popular for jet reconstructions for various reasons. Jet reconstruction
techniques in hadron colliders are ought to be different than what
is followed in $e^+e^-$ experiments. In hadron colliders, it is not possible to measure the total
energy (used in {\tt JADE} algorithm~\cite{jade}) accurately because of loss along the beam pipe,
and the QCD divergences occur not only between the outgoing particles,
but also between an incoming and outgoing particle as well.
The $k_T$ algorithm
~\cite{Catani:1992zp} is devoid of all these difficulties and suitably defined for
hadron collider environment. It is advantageous to define jet algorithms in terms of variables,
which are invariant under longitudinal boost, so that it gives the same output in laboratory
and center of mass (COM) frame.Interestingly the 
longitudinal $k_T$
algorithm~\cite{Catani:1992zp,fastjet} exactly delivers this boost invariant formalism.
The formulation of $k_T$ algorithm can be described very briefly with the general 'distance' formula as,
\begin{eqnarray}
 d_{ij}&=&min(p_{Ti}^{2m},p_{Tj}^{2m})\frac{\Delta R_{ij}^{2}}{R^{2}},\nonumber  \\
\Delta R_{ij}&=&\sqrt{(y_{i}-y_{j})^{2} + (\phi_{i}-\phi_{j})^{2} },\nonumber \\
d_{iB}&=&p_{Ti}^{2}\nonumber, 
\label{eqn:kt}
\end{eqnarray} 
where $d_{ij}$ is the distance between two particles and $d_{iB}$
is the particle-beam distance, and R plays the same role as the cone
radius, eq.~\ref{eq:dR}. Here all the quantities related to the kinematics
of jet viz. $\Delta y~=~y_{i}-y_{j}$, $\phi_{i}$, $p_{Ti}^{2}$, and
hence $ d_{ij}$ are invariant under longitudinal boosts. Here $y$ refers to rapidity of a 
 particle, defined to be $y~=~\frac{1}{2}ln(\frac{E+p_{z}}{E-p_{z}})$.
 For $k_T$ algorithm,
the parameter $m$ is set equal to unity. The workflow of the algorithm is as following
~\cite{Catani:1992zp,fastjet,PhysRevD.48.3160}: 
\\
1. For all stable particles in an event, $ d_{ij}$ and $d_{iB}$ are
evaluated.
\\
2. The minimum of $ d_{ij}$ and $d_{iB}$ are checked. If $ d_{ij}$ is
smaller among the two, then the two particles are combined
to form a single new particle and calculation restarts from step 1.
The momenta of the new particle is the sum of four momenta of individual
particles.
\\
3. If $d_{iB}$ is smaller among the two, then the particle $i$ is
declared to be the final state jet. It is removed from the list
of particles and the algorithm restarts from step 1.
\\
4. Steps (1) - (3) continues until no particles are left in the event. \newline

As stated earlier,
 The number of final state particles clustered into jets solely
depend on the jet radius R. However, a drawback of this algorithm is
that arbitrarily soft particles enter into the jet radius leading to a possible
contamination of jet energy. Hence while using this algorithm one has to
select particles with a minimum $p_{T}$ threshold. Evidently, this algorithm is free
from any kind of divergence issues. Incidentally the algorithm is computationally
slow, time taken is order of $N^{3}$ where N is the number of initial 
particles and also produces irregular shaped jets~\cite{Salam:2009jx}.
The another IRC safe jet algorithm is the C/A algorithm
~\cite{Dokshitzer:1997in}  which is basically
an angular ordered jets, defined with $m$ = 0 in eq.~\ref{eqn:kt}. In hadron colliders, it works like $k_T$
algorithm, i.e it involves a R cut, instead of a angular cut. It
calculates $\Delta R_{ij}$, eq.~\ref{eq:dR} and continues to calculate it until
all objects are separated by a $\Delta R_{ij}>$ R cut. Eventually all
the final objects are selected as jets. This C/A algorithm also suffered
by practical problems as $k_T$ algorithm, as discussed before.
The anti-$k_T$ algorithm which is the amalgamation of $k_T$
and C/A algorithm is comparatively free from all problems including IRC singularity. \newline 

The formulation of anti-$k_T$
can be obtained by setting $m$ = -1 in eq.~\ref{eqn:kt}, leading the
name anti-$k_T$. Clearly, this algorithm, while proceeding to reconstruct,
gradually include only the hard particles resulting a growth of jets
outwards from the jet axis~\cite{antikt}.
There are certain key features of this
algorithm which makes it very useful. If there are several soft particles
of transverse momenta $p_{Ti}$ in the vicinity of a hard particle with
momentum $p_{T1}$, then the distance between the hard particle and any
other soft particle is dominantly determined by the momentum of the hard
particle(since $p_{T1}~\gg~p_{Ti}$). In the absence of any other hard particles
within a distance $2R$, then all the soft particles will be clustered
with the hard particle to form an exact conical jet of radius $R$. On the other hand,
if there are two hard particles with momenta $p_{T1}$, $p_{T2}$ then
there will be two jets, none of which will be exactly conical. The
boundary $b$ between these two jets is defined by,
$\Delta R_{1b}/p_{T1}~=~\Delta R_{2b}/p_{T2}$.
The essential feature is that soft particles don't effect the jet
boundary like the hard particles. Hence this algorithm is insensitive
to soft emission which makes it IRC safe. In all present hadron collider experiment, particularly at the LHC,
anti-$k_{T}$ algorithm is widely used for jet reconstruction due to its robustness.

\section{Jet reconstruction in CMS}
\subsection{Apparatus}
Compact Muon Solenoid (CMS) is one of the two multipurpose detector in LHC, the other one
is ATLAS.
 The CMS detector is cylindrical in shape and the CMS coordinate system has its origin at the center of the detector,
 with the $z$-axis pointing along the direction of the
 counterclockwise LHC beam. The central feature of the CMS detector
 is a superconducting solenoid, of 6 meter internal diameter, that produces an axial magnetic field of 3.8T.
 Within the field volume there are the silicon pixel and strip tracker, a lead-tungstate crystal
 electromagnetic calorimeter (ECAL) and a brass/scintillator hadronic calorimeter (HCAL). Outside the field volume,
 in the forward region ($3 < |\eta| < 5$), there is an iron/quartz-fiber hadronic calorimeter.
 Muons are measured in gas detectors embedded in the steel return yoke outside the solenoid, in the pseudo-rapidity
 range $|\eta| < 2.4$.
 
 In the region $|\eta| < 1.74$, the HCAL cells have widths of
0.087 in pseudo-rapidity and 0.087 in azimuth ($\phi$). In the
$\eta$-$\phi$ plane, and for $|\eta|< 1.48$, the HCAL cells map on
to $5 \times 5$ ECAL crystals arrays to form calorimeter towers
projecting radially outwards from close to the nominal interaction
point. At larger values of $|\eta|$, the size of the towers
increases and the matching ECAL arrays contain fewer crystals. Within
each tower, the energy deposits in ECAL and HCAL cells are summed to
define the calorimeter tower energies, subsequently used to provide
the energies and directions of hadronic jets.

At 7 TeV energy the energy resolution for photons with $E_T \approx 60$ GeV varies
between 1.1\% and 2.5\% over the solid angle of the ECAL barrel, and
from 2.2\% to 5\% in the endcaps. The HCAL, when combined with the
ECAL, measures jets with a resolution $\Delta E/E \approx 100\% /
\sqrt{E\,[\rm GeV]} \oplus 5\%$.
 A more detailed description of the CMS experiment can be found elsewhere~\cite{CMSDETECTOR}.

\subsection{Jet Reconsruction}
\label{sec:jetreco} 
In experiment, the ingredients to the jet reconstruction algorithm are the momenta of final state stable 
particles or energies of the calorimeter towers. Jets, in particular measured by 
detectors are broadly classified into different categories based on the type of inputs
passed to the jet reconstruction algorithm.  
For example, in CMS, based on the sub-detector
inputs, type of jets are calorimeter (Calo) jets, jet-plus-track (JPT) jets 
and particle flow (PF) jets~\cite{CMS-PAS-PFT-09-001}. 
Calo jets~\cite{CMS_PAS_JME_07_003,CMS-PAS-JME-10-003} are
reconstructed from the energy deposits in
both electromagnetic and hadronic calorimeter cells. The cell energies are then
combined to form energy towers which are used as a input to jet algorithm to reconstruct
Calo jets. The JPT~\cite{CMS_PAS_JME_09_002} jets are
reconstructed from calorimeter deposits also but corrected for the
momentum of charged particles associated with track informations from tracker.
The PF~\cite{CMS-PAS-PFT-09-001} jets are reconstructed from the informations
of each individual visible particles.
Each visible particles are reconstructed combining informations from all 
sub-components of the detector. More precisely, the charged particles, electrons, 
muons and charged hadrons are reconstructed from the tracks in the 
tracker where as photons and neutral hadrons are reconstructed from 
Electromagnetic calorimeter (ECAL)
and hadron calorimeter (HCAL). The energy of electrons are measured
combining the measurement of tracks and the corresponding energy deposits
in ECAL. The energy of muons are estimated from the curvature of the tracks in tracker and muon chamber.
Inputs from tracker and corresponding energy deposits in ECAL and HCAL are used to measure the 
energy of charged hadrons, where as the informations from the calorimeters
are only considered to construct energy of neutral hadrons. The energy of photons is obtained from 
ECAL directly. In order to obtain a very accurate estimation of 
energy-momentum of jets, it is mandatory to reconstruct stable 
individual particles, the constituents of jets, with a 
very good resolution. The PF algorithm delivers it by measuring 
the energy of all the stable particles, particularly charged hadrons 
and photons which constitute almost ~85\% of jets, very precisely using 
tracking detectors with high resolution and ECAL with high granularity. 
For example, particles like electron, muon and
$\pi^{+},\pi^{-},\pi^{0},K^{+},K^{-},K^{0}_{L},\gamma$
are reconstructed with high momentum resolution using the PF technique.
Finally, in CMS the jet reconstruction technique is applied on the
collection of particles~\cite{CMS-PAS-PFT-09-001} interfacing the 
 FastJet~\cite{fastjet} package and using anti-$k_{T}$ algorithm. \newline

It is to be remembered that the measured detector jet energies are not expected to
be the same as the corresponding true particle level jet due to the non-linear
response of the detector. Hence, in order to achieve a correct 
estimation of jet energy, on average, the detector jet energy has to be corrected. 
In CMS, this correction factor is obtained by factorizing it into different levels e.g.
 off-set correction, the MC calibration factor, relative and 
residual correction. The off-set correction is mainly to eliminate the
effects of instrumental noise and pile-up (PU) effects, the MC correction 
is due to the non-uniformity in $\eta$ and $p_T$, where as relative
correction is for the $\eta$ and  residual correction is finally
to take care the left over differences between data and simulation~\cite{1748-0221-6-11-P11002}. \newline

The charged hadrons and photons have much better resolution
as they are reconstructed from tracker and ECAL respectively. 
The
correction in jets is mainly required for the neutral hadron
components, for which the detector response is non uniform. 
The $\eta$ dependent correction factor (L2 factor) is determined 
from dijet balance and where as the $p_T$ dependent factor (L3 factor) is obtained
from $\gamma+$jet and $Z+$jet events. 
These correction factors are determined event by event simulated by PYTHIA
\cite{pythia} and processed through the CMS detector simulation based on
GEANT4~\cite{geant4}. An additional offset correction is also taken into 
account due to the excess energy coming from particles which originate 
from the secondary vertices in the same or neighboring branch
crossings (L1 factor). It is determined by computing the energy density ($\rho$) times the jet area ($A$). 
The effects of pile-up are most dominant on low $p_{T}$ and is 
negligible for high $p_{T}$($\ge 200~GeV$) jets. 
Eventually the total
correction factor which is approximately $\sim$1.2 at lower $p_{T}$ range 
and reduces to $\sim$1 at higher range of $p_{T}$ of jets. 
The jet $p_{T}$ resolution is of the order of $10\%$ at $p_{T}=100~GeV$ where
as the dijet mass ($M_{jj}$) resolution varies from $7\%$ to $3\%$ for the range of 
$M_{jj}$ from $0.2~TeV$ to $3~TeV$~\cite{QCD-11-004}. \newline

\begin{figure}
\includegraphics[scale=0.35]{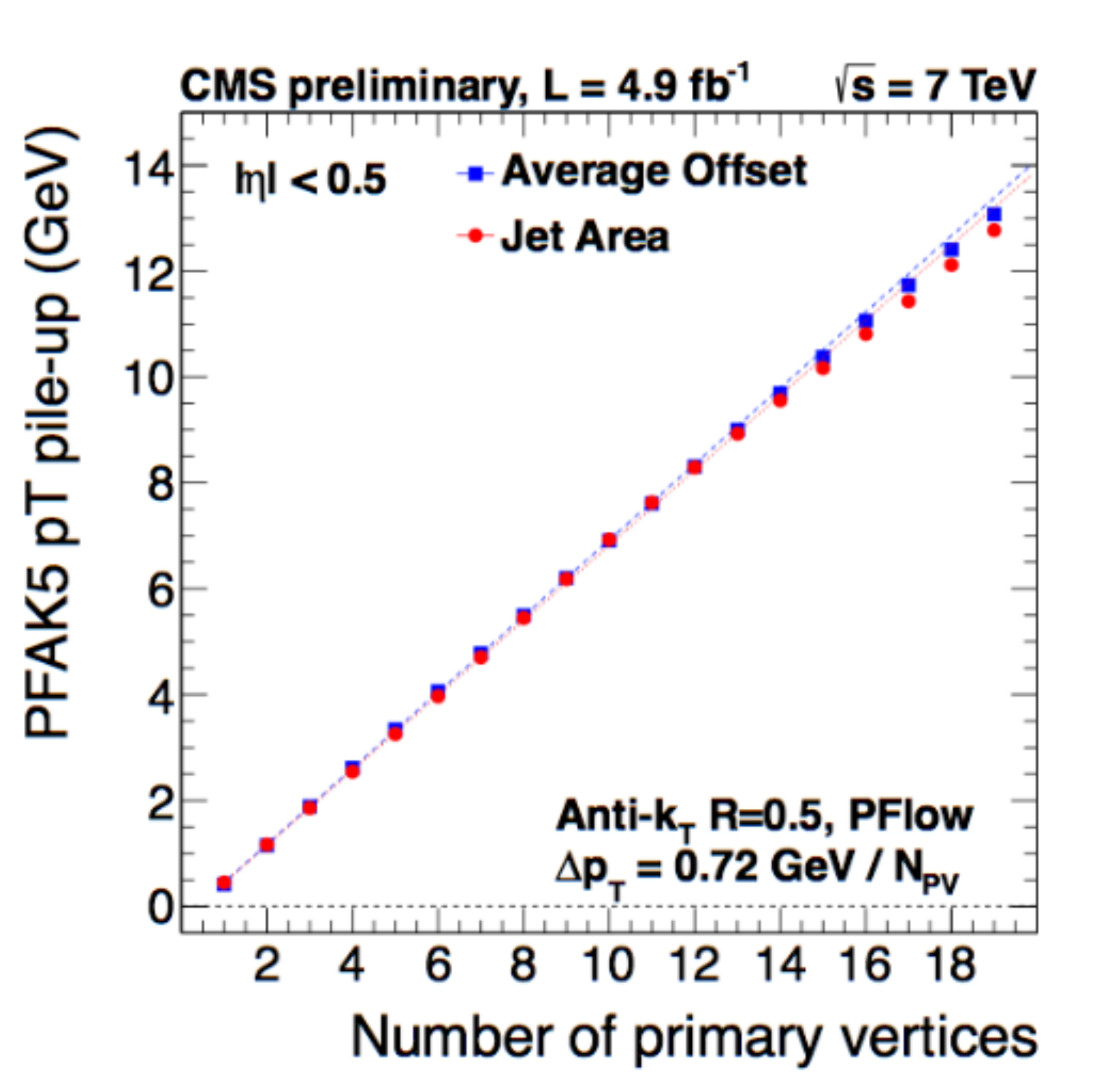}
\includegraphics[scale=0.35]{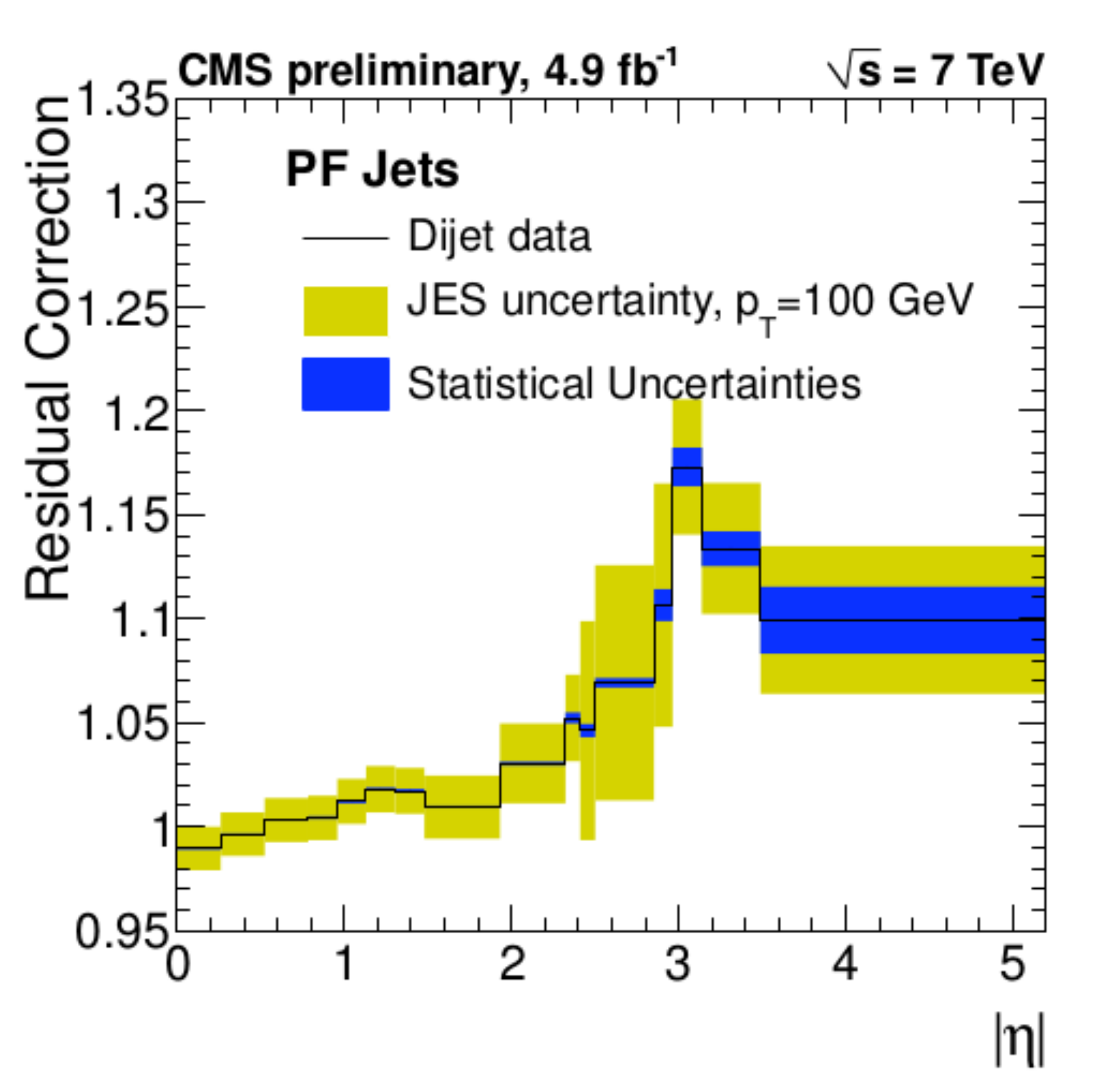}
\caption{PU Energy vs number of primary vertex and Relative Factor vs $|\eta|$ } 
\label{fig:jec}
\end{figure}

\begin{figure}
\hspace{3cm}
\includegraphics[scale=0.5]{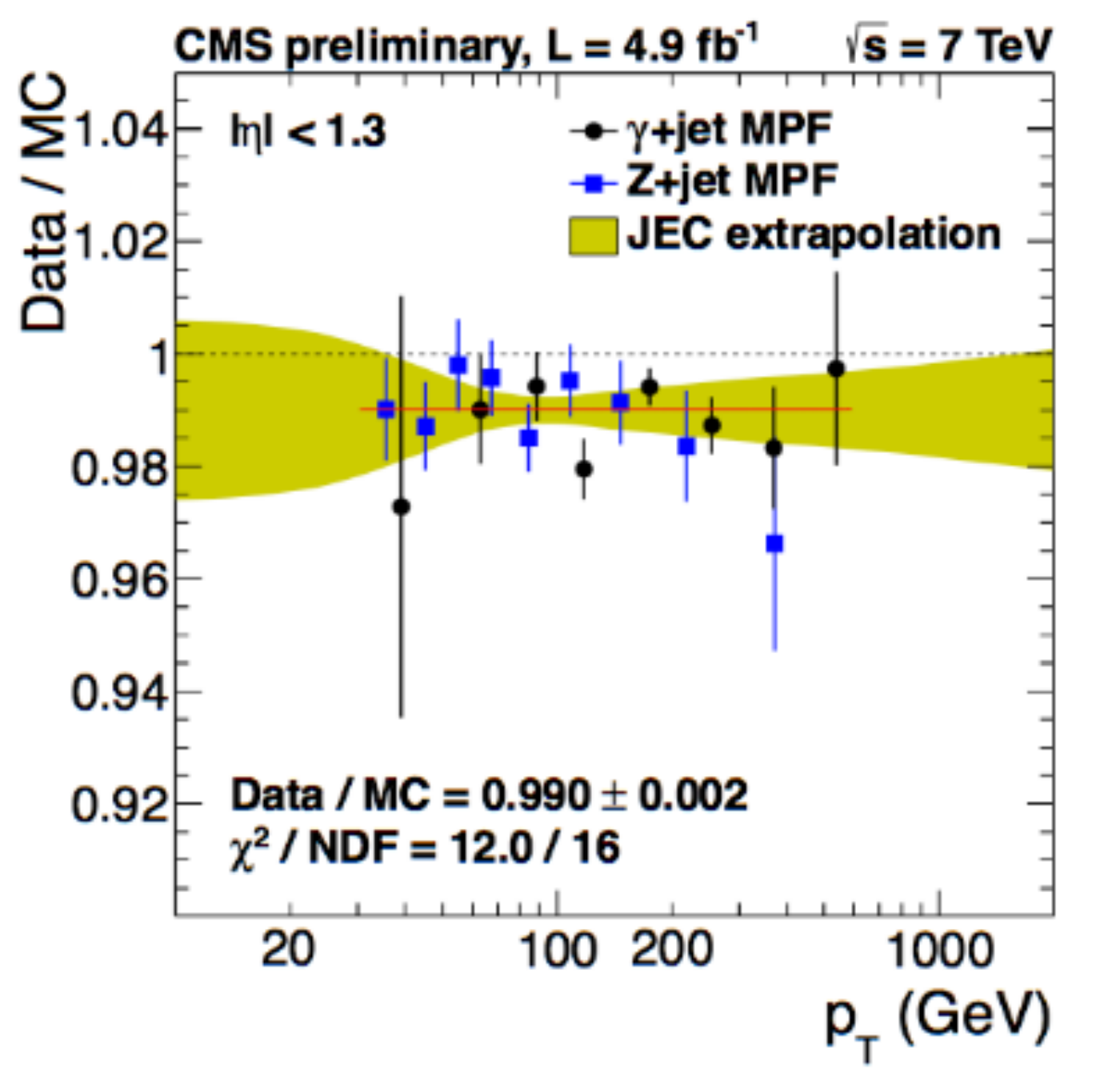}
\caption{$p_{T}$ dependent correction factor}
\label{fig:jecl3}
\end{figure}

Fig.~\ref{fig:jec} displays the level of jet corrections required to be taken into account corresponding to 
CMS detector. It shows that the amount of PU energy increases with the 
number of good primary vertices in an event and it is 0.72 GeV per
primary vertex. The plot on left panel shows the offset correction as function of good reconstructed primary
vertex. The plot on  right panel shows the amount of $\eta$ dependent correction for its wide range
along with other uncertainties viz. jet energy scale (JES) and statistical uncertainty~\cite{1748-0221-6-11-P11002}.\\
Finally, Fig.~\ref{fig:jecl3} shows the ratio between data and MC after $p_{T}$ dependent correction is applied. On an 
average the $1\%$ mismatch between data and MC is applied as the residual correction~\cite{1748-0221-6-11-P11002}. \\

\section{Jet Cross-Section Measurement}
\begin{center}
\begin{figure}
\includegraphics[scale=0.6]{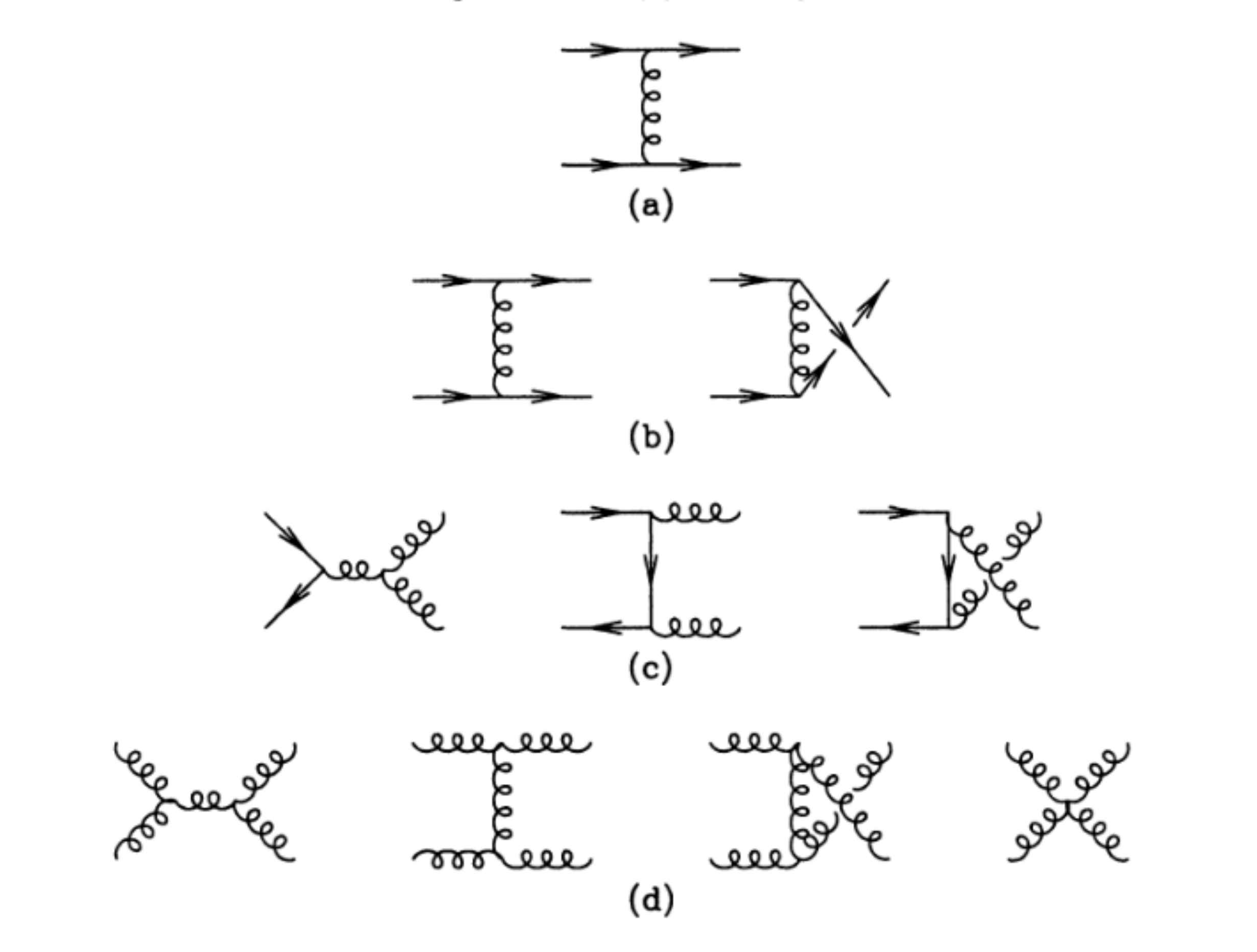}
\caption{Feynman diagrams for tree level processes contributing in jet production at hadron colliders}
\label{fig:feyn_diagram}
\end{figure}
\end{center}

In proton-proton collision, the total 
scattering cross section is computed convoluting the parton distribution 
function (PDF) of each incoming parton from each proton with the corresponding 
partonic level cross section. At leading order (LO), the jets are produced via the
subprocesses, $q q \to q q $, $q\bar q \to q \bar q$, $gg \to gg$, $qg \to qg $,
$gg\to q\bar q$ and $q \bar q \to gg$ where the leading partonic level cross
section turns out to be $gg \to gg$ because of the large color factors. 
In Fig.~\ref{fig:feyn_diagram}, we show the Feynman diagrams for these sub processes.
Note that, relative contributions due to these sub channels  
to the total cross section is a combine effect of initial
PDF and the magnitude of partonic level cross section.  
If the two incoming protons carry four
momenta $P_{1}$, $P_{2}$, then the differential total cross section
is given by,
\br
 d\sigma(P_{1}, P_{2})= \sum_{a,b} \int \int dx_{1}
dx_{2}~ f_{a}(x_{1},\mu_{F}^{2})~ f_{b}(x_{2},\mu_{F}^{2})~
d\hat{\sigma}
(p_{1}, p_{2}, \mu_{R}^{2}, \alpha_{S}(\mu_{R}))
\er

Here $p_{1}$, $p_{2}$ are the four momenta of the two incoming partons
which take part in the hard interaction and $x_i$'s are the Bjorken
variables defined to be the momenta fractions $x_{i}~=~p_{i}/P_{i}$, 
for $i=1,2$. The sum over indices
$a$ and $b$ run over different flavors of incoming partons.
The scales $\mu_{F}$, $\mu_{R}$  denote the factorization and renormalization
scales respectively, $\alpha_{S}(\mu_{R})$ is the  strong coupling
constant evaluated at the scale $\mu_{R}$.
The $\hat\sigma$ is the partonic level cross section 
calculated using the principles of pQCD.
The LO QCD cross section is enormous which is of the order of
$\sim \alpha_s^{2}$ is found to be $\sim$ $10^{8}$pb for $p_{T} \ge $10 GeV
 and comes down to $10^{5}$pb for $p_{T} \ge $100 GeV for $\sqrt{s}=$7 TeV. Since it is predominantly a QCD process and 
mediated by gluon, an uncertainty due to the choice of scales and also
PDF is expected to be very large($\sim$100\%). Therefore, in order to 
obtain a reliable estimate of the jet cross section, one needs to consider next to leading order (NLO) terms
in perturbation theory. Currently the NLO jet cross section is computed maximum up to 5 jets final state~\cite{Bern:2013gka}.
 In this study 
the NLO calculation for jet cross section is performed using NLOJet++\cite{Nagy:2003tz}
package.


\subsection{Jet and Event Selection}
\begin{figure}
\includegraphics[scale=0.35]{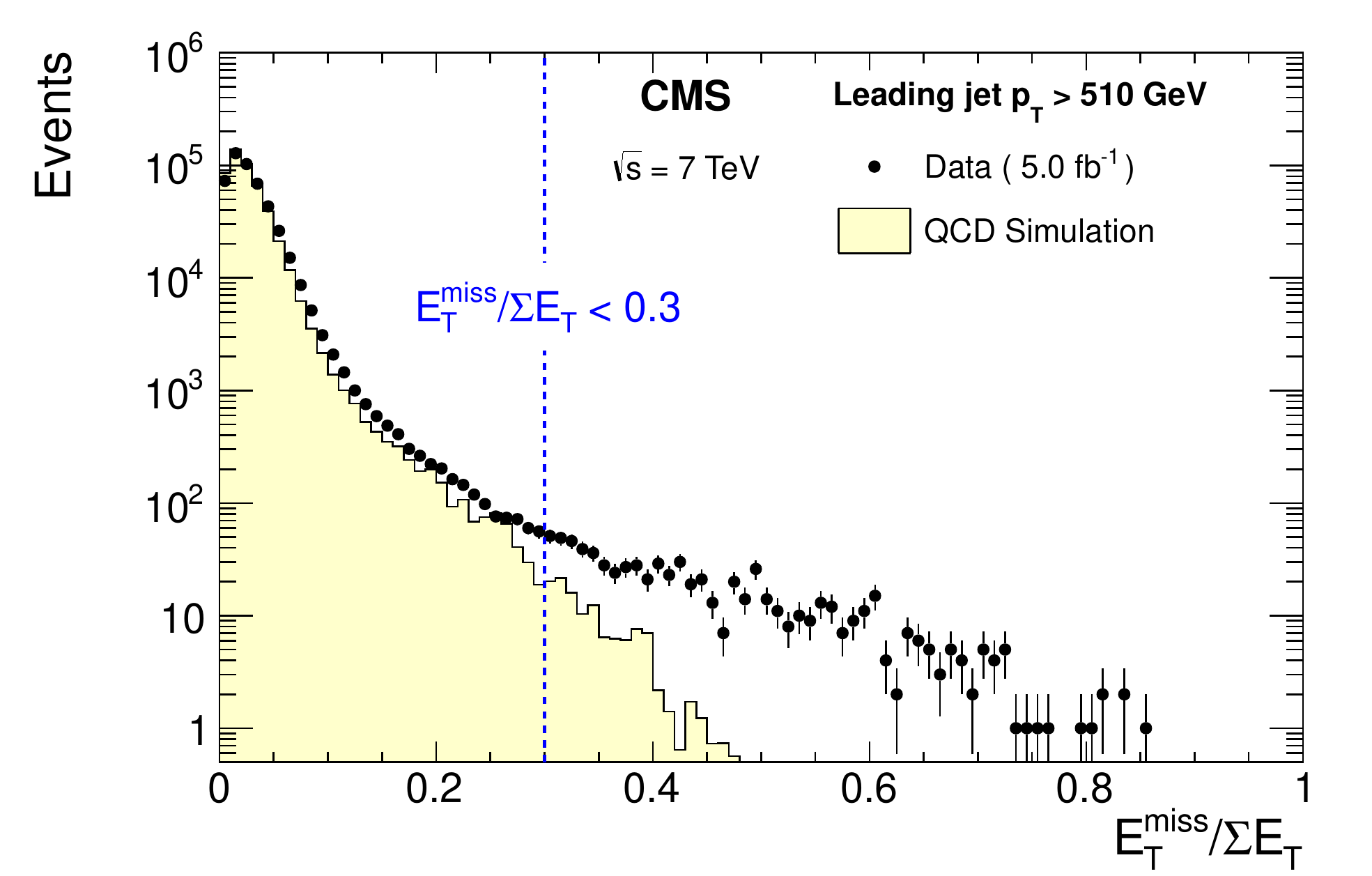}
\includegraphics[scale=0.35]{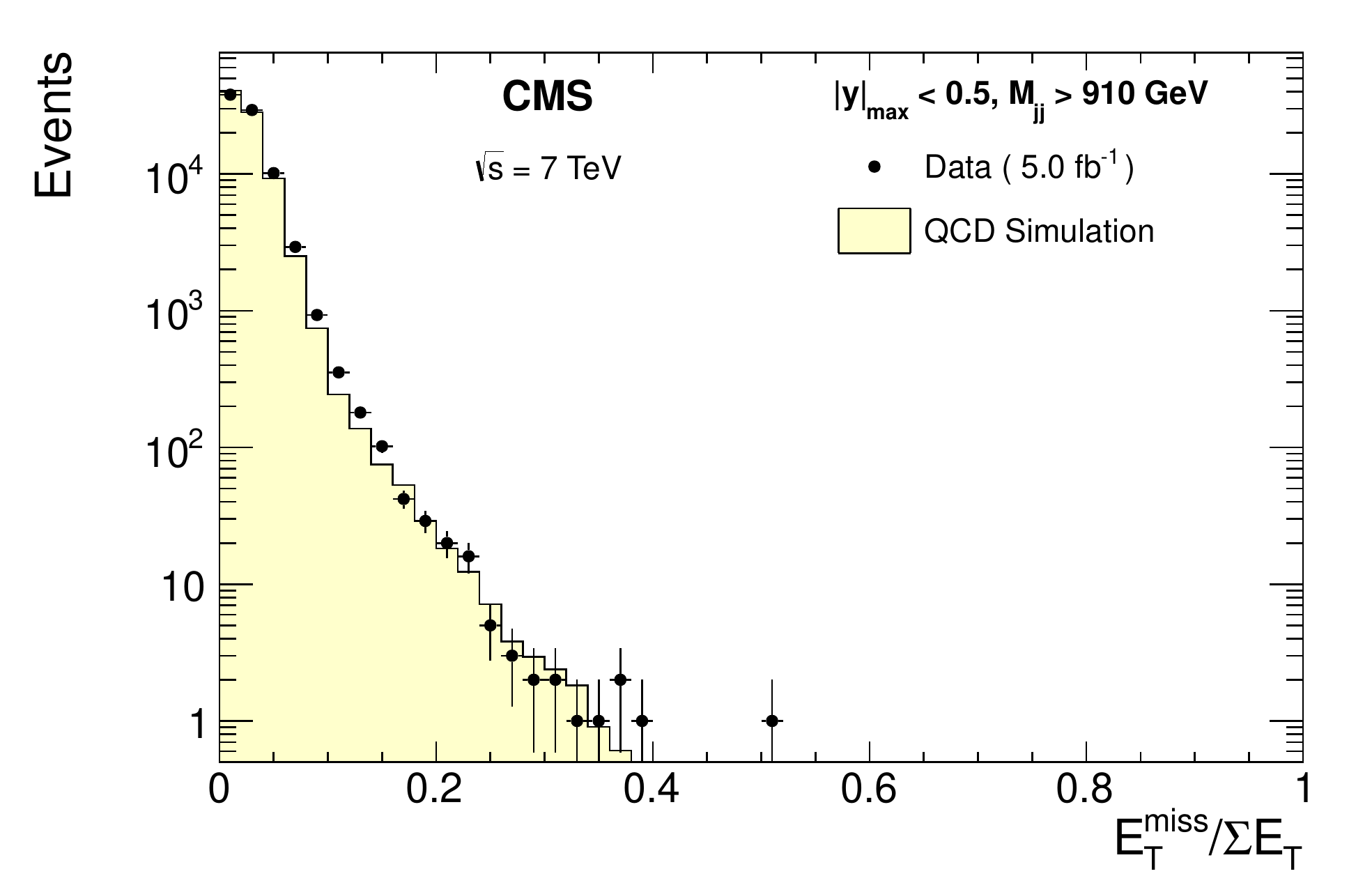}
\caption{ $|\vec{\ensuremath{{\not\! E}_{T}}|}/ \sum E_T$ distribution with
 inclusive jet and dijet event selection criteria}
\label{fig:met_sumet}
\end{figure}

As mentioned in the previous section, in the CMS experiment jets are reconstructed using 
anti-$k_T$~\cite{antikt} algorithm built in FastJet package~\cite{fastjet} with the 
size parameter R=0.7. The choice of larger value for R  
allows to cluster more hard scattered partons and hence the jet energy
and dijet mass resolution is increased compared to smaller 
value of $R$. A jet with energy $E$ and momentum components 
$\vec{p}~=~(p_{x},p_{y},p_{z}) $ will have 
transverse momentum $p_{T}~=~\sqrt{p_{x}^{2}+p_{y}^{2}}$.
In this measurement high quality of events are ensured by imposing certain selection criteria. 
For example, the event should have a good reconstructed primary vertex to which 
at least four well reconstructed tracks are associated. The vertex should be
within a distance along the z axis (original beam direction) from the center of the detector ($|z|<24$cm)
and in the x-y plane (transverse to z axis) it can be shifted at most 2cm ($\sqrt{x^{2}+y^{2}}<2$cm). Indeed
any pure QCD event is expected to have a negligible
missing transverse energy (MET), as defined, 
 $\vec{\ensuremath{{\not\! E}_{T}}}~=~ -\sum_{i}(E_{i}~\sin \theta_{i} ~\cos \phi_{i} ~\hat{x}
~+~E_{i}~\sin \theta_{i} ~\sin \phi_{i} ~\hat{y})$, the
summation over $i$ runs over all the reconstructed particles in the event. 
$E_{i}$ is the energy of the $i$ th particle,
and $\theta_{i}$,$\phi_{i}$ are the polar and azimuthal angle of the 
corresponding particles measured with respect to
$\hat{z}$, the initial beam direction, $\hat{x},\hat{y}$ are the 
unit vectors along $x$ and $y$ axis respectively. 
Naturally, the ratio, $|\vec{\ensuremath{{\not\! E}_{T}}|}/\sum_{i}E_{T}^{i}$ is a good discriminator to 
isolate pure QCD events as shown in Fig.~\ref{fig:met_sumet}, where it is shown for both data and MC
corresponding to inclusive jet(left) and dijet events(right)~\cite{QCD-11-004}.
We choose to apply an upper cut 0.3 to select genuine QCD events rejecting
any contribution due to noise of any miss calibration of the detector. A long tail of
 $|\vec{\ensuremath{{\not\! E}_{T}}|}/\sum_{i}E_{T}^{i}$ beyond 0.4 may be due to events from $Z$+jets,
 where $Z\rightarrow \nu \bar{\nu}$ process leads to a high missing energy.
Once a good event is selected then jets are formed by clustering the 
stable particles reconstructed by PF technique in the event. 
Reconstructed jets are checked with the jet identification criteria 
to ensure good quality of jets originating from hard scattered partons 
and not due to some detector level noise. The tight jet 
identification criteria applied in CMS is the following~\cite{QCD-11-004} \\
\textbullet At least one PF particle.\\
\textbullet Charged hadron energy fraction and multiplicity $\geq 0$ 
in the region $|\eta| \leq 2.4$.
\textbullet Neutral hadron energy fraction $\leq 0.9$. \\
\textbullet Photon energy fraction $\leq 0.99$ . \\
\textbullet Muon energy fraction $\leq 0.9$. \\
\textbullet Electron energy fraction $\leq 0.9$. \newline
In addition, as discussed in previous section, energy correction is required 
on the measured jet energy to account for the non-uniform and non-linear response of 
CMS calorimetric system. After the jet energy correction is
applied, the measured momenta are corrected to the particle level.
The events, produced after proton-proton collision, are filtered by a trigger system and then stored.
In CMS, trigger is a two tier system, Level-1~\cite{L1TRIG} and High Level Trigger (HLT)~\cite{HLT}.  
The former one is mainly a hardware based trigger whereas the later one is a software based trigger.\newline

The data which is analyzed for this measurement collected by six HLT of 
$p_T$ thresholds 60, 110, 190, 240, 370 GeV. Events consisting at least one jet
with corrected jet $p_T$ greater than the trigger threshold are stored.
The on line triggered jets are calorimeter jets with worse resolution 
where as off-line jets are constructed by particle flow technique and has better resolution compared to calorimeter jets.
The triggers with lower $p_{T}$ thresholds have high prescale factors 
to fit the trigger bandwidth due to the high QCD event rates and hence have low 
effective luminosity. In
Table~\ref{table:efflumi} we show the individual HLT paths 
with the corresponding effective integrated luminosities. 
In order to achieve full trigger efficiency, we apply a off-line cuts 
on PF jet $p_{T} \ge $ 114,
196, 300, 362 and 507 GeV for each HLT triggers respectively.  
The inclusive jet cross section
measurement is carried out up to a rapidity 2.5 in an equal bins 
of $\Delta y=0.5$. For the dijet mass measurement, at least two jets 
with momenta $p_{T1} \ge 60$ GeV and
$p_{T2} \ge 30$ GeV are required in the event. The cross section is measured in the bins of 
maximum rapidity $y_{max}=max(|y_{1}|,|y_{2}|)$. Low $y_{max}$
value probe the large angle of scattering in the s-channel while 
higher value of $y_{max}$ probe small-angle scattering in t-channel.

\begin{table}
 \begin{center}
 \begin{tabular}{|c|c|c|c|c|c|}
 \hline
  HLT $p_{T}$ ($GeV$) & 60   & 110 & 190 & 240 & 370 \\
 \hline
  $L_{eff}~(pb^{-1})$   & 0.4  & 7.3 & 152 & 512 & 4980 \\
 \hline
 \end{tabular}
 \caption{HLT Triggers And Effective Luminosities}
 \label{table:efflumi}
 \end{center}
 \end{table}

\subsection{$p_{T}$ and $M_{jj}$ Measurement}
\begin{figure}
\includegraphics[scale=0.35]{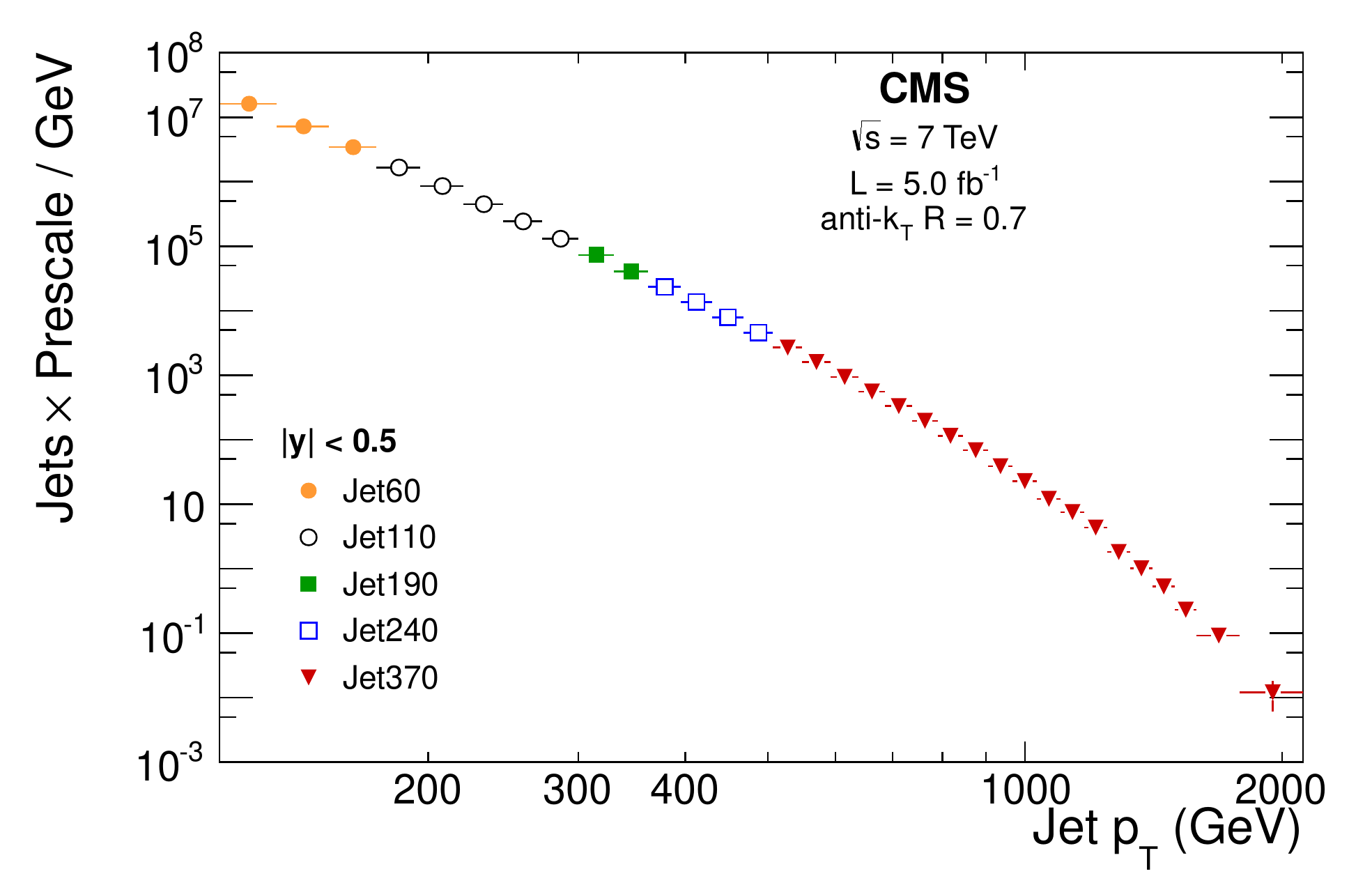}
\includegraphics[scale=0.35]{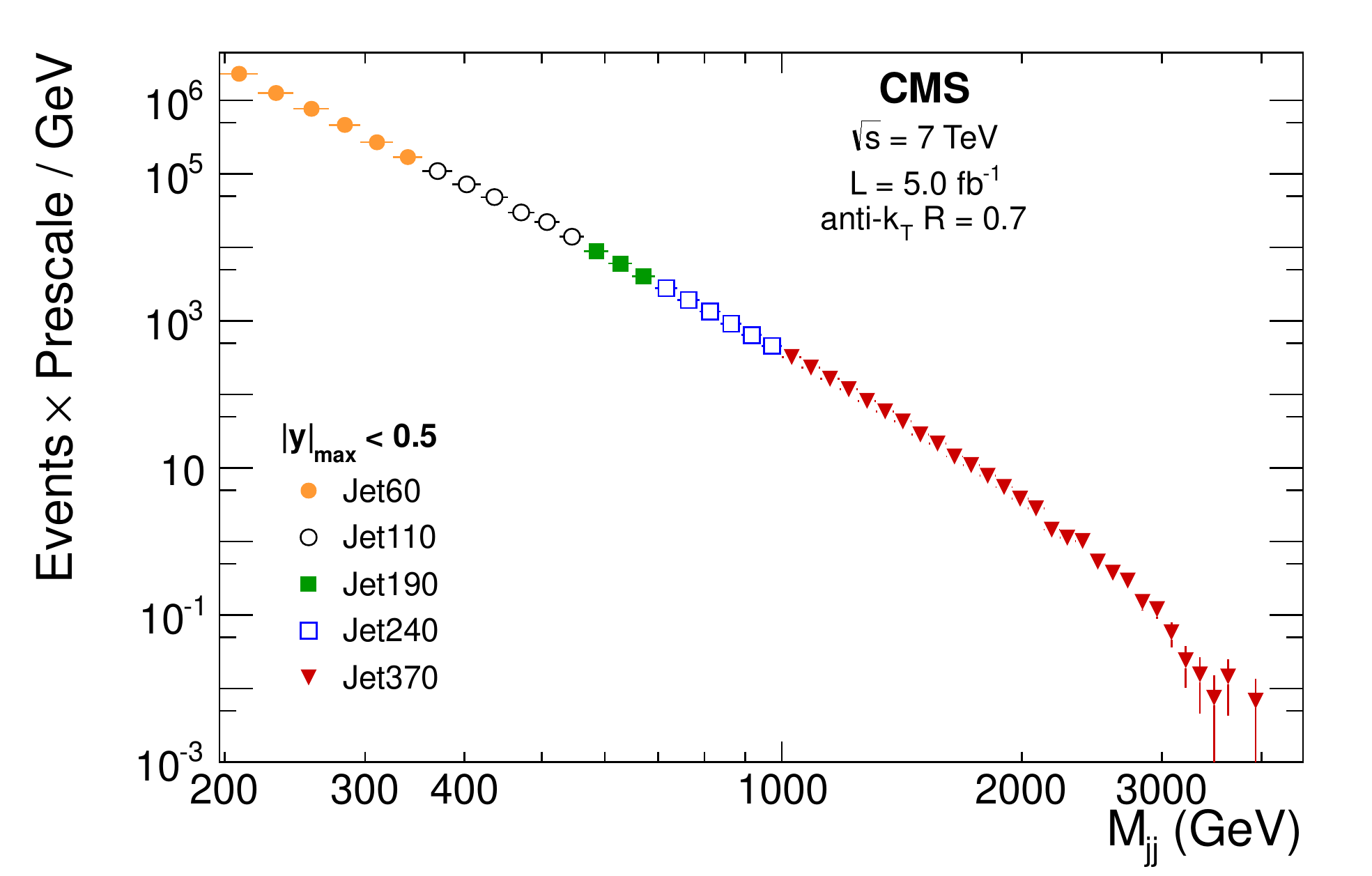} \\
\caption{Measured jet yield scaled by the trigger prescale factor for inclusive jet production on left pannel
and on right panel the same distributions shown for dijet mass distribution for the
central rapidity bin $|y|<0.5$. }
\label{fig:yield}
\end{figure}

The pure QCD events, in collected data sample, are isolated requiring jets and events should pass certain selection criteria
as described in the previous section.
 The distribution of transverse momentum of jets, 
$p_{T}$ are obtained dividing the entire $p_T$ range into 21 bins
for six rapidity intervals with $\Delta |y|$=0.5 and also similarly for  
dijet invariant mass ($M_{jj}$) spectra are obtained.
The entire range of $p_T$ distribution is obtained  corresponding to 
each five different HLT paths.  
The reconstruction of each segment of the spectrum is obtained 
only by one trigger path avoiding double counting of jets. Eventually, the 
measured yields are transformed to double-differential 
inclusive-jet cross sections as:

\br
\label{eq:inclusive}
\frac{\text{d}^2\sigma}{\text{d}p_{T}\text{d}|y|}=\frac{1}
{\epsilon\mathcal{L}}\frac{N}{\Delta p_{T} \Delta |y|}\text{,}
\er
 where $N$ is the number of jets in the corresponding $p_{T}$ bin, 
$\mathcal{L}$ is the effective integrated luminosity of the data sample
 taking into account the trigger prescales.
Here $\epsilon$ is the product of the trigger and jet selection 
efficiencies, $\Delta p_{T}$ and $\Delta |y|$ are the 
corresponding bin widths. Notice that the bin-width increases progressively, 
proportional to $p_{T}$ resolution. 
By a similar fashion, the double differential cross section for di-jet mass
distribution is obtained as,
\br
\label{eq:dijet}
\frac{d^2 \sigma}{dM_{jj}d|y|} = \frac{1}{\epsilon \mathcal{L}}
\frac{N}{\Delta M_{jj} \Delta |y|}\text{,} 
\er
where the symbols represent identical meaning as the previous 
equation, eq.\ref{eq:inclusive}, $\Delta M_{jj}$ is the dijet 
invariant mass bin width
which increases progressively as well, equal or larger 
than $\Delta M_{jj}$ resolution.\\
Fig.~\ref{fig:yield} shows the measured differential cross-section for various $p_{T}$ (left)
and $M_{jj}$ (right) as directly
measured from data for the central rapidity bin. Each region of $p_{T}$ or $M_{jj}$ is constructed using
the particular HLT. Notice that, for $\sqrt{s}~=~7$ TeV, $p_{T}$ extends up to $\sim$ 2 TeV and $M_{jj}$ extends
up to $\sim$ 4 TeV. 

The statistical uncertainty in the number of jets in a bin is \\
$e_{stat}~=~\sqrt{(4-3f)/(2-f)}.\sqrt{N_{jets}}$, where $f~=~N_{1}/N_{ev}$ is the 
fraction of events that contribute one jet in the given bin. The formula
is valid under the assumption that the number of events that contribute more than two jets in
each bin is negligible, which has been verified for the current measurement.

\begin{figure}
\includegraphics[scale=0.35]{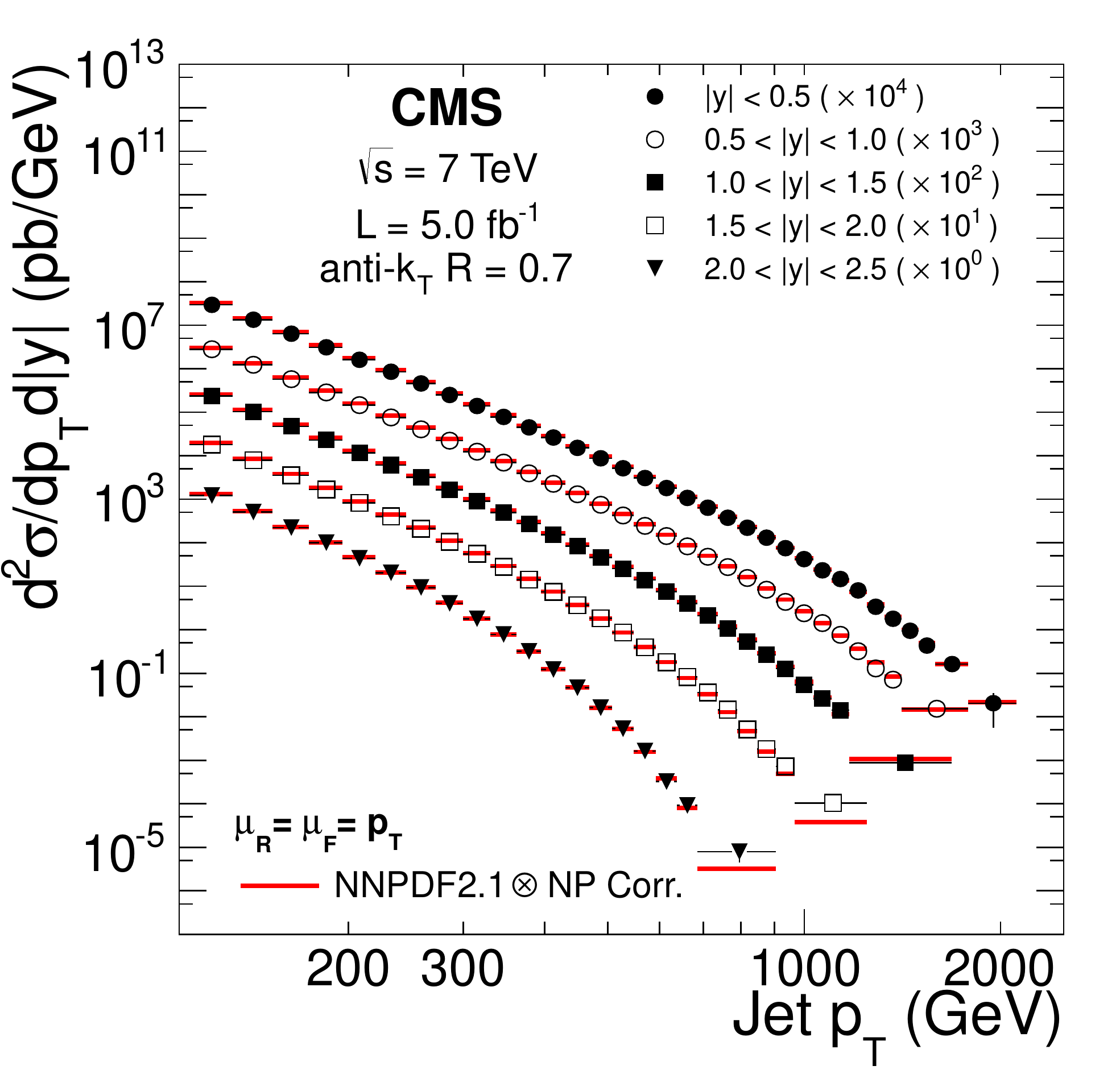}
\includegraphics[scale=0.35]{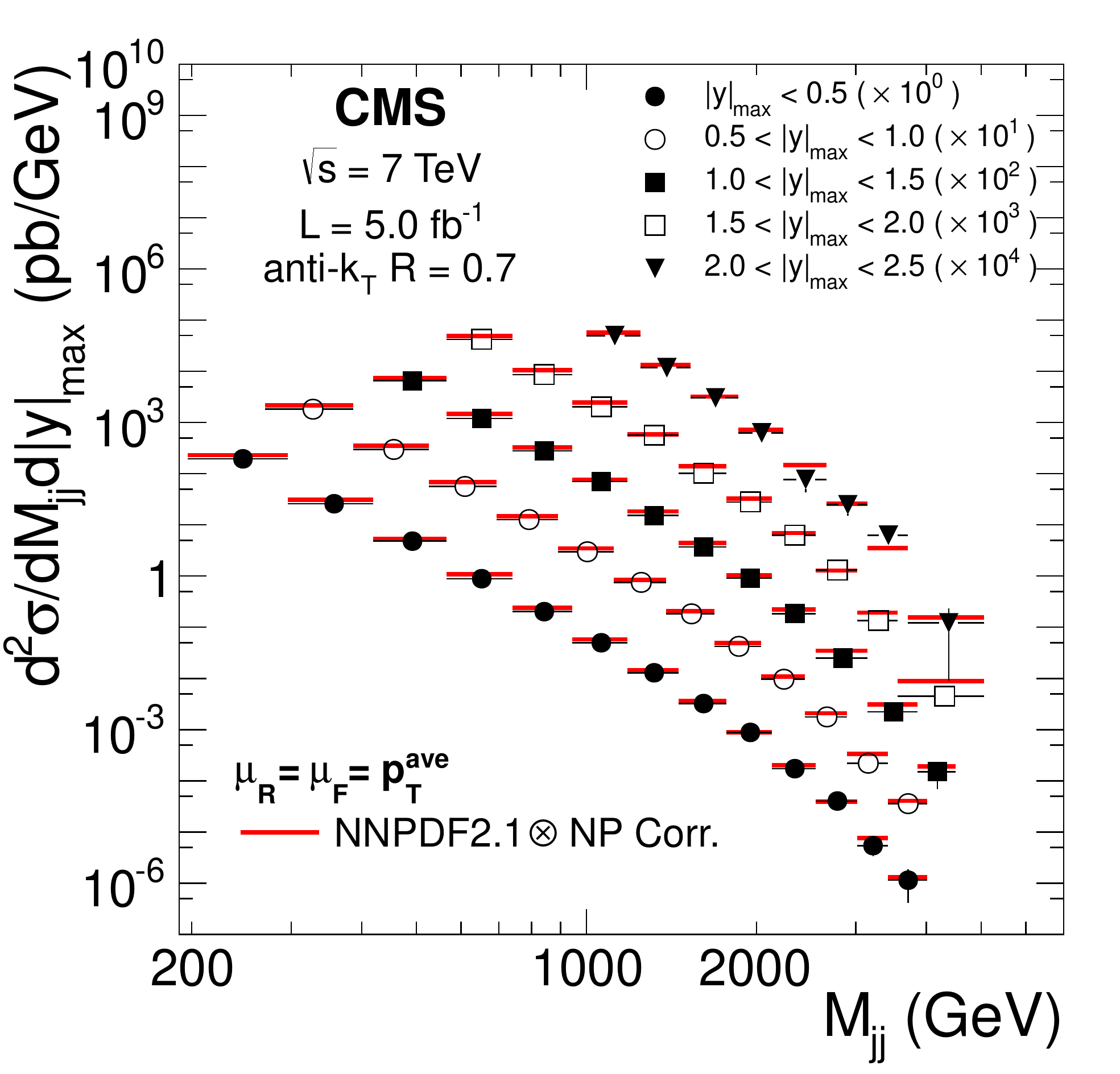} \\
\caption{ $\rm NLO \times \rm NP$ theory prediction
(with NNPDF2.1 PDF set) is compared with the measured spectrum
 up to rapidity $|y| = 2.5$ at 
an interval of 0.5 }
\label{fig:spectrum}
\end{figure}

In Fig.~\ref{fig:spectrum}, we present the inclusive jet $p_T$
spectrum on the left and the di-jet mass spectrum on the right 
for various $y$ bins corresponding to $\lumi=5 \invfb$. 
The distributions are scaled up by a factor for better presentation 
as shown in the plot. 
The theory predictions are estimated, as mentioned before, by using NLOJet++~\cite{Nagy:2003tz} and NNPDF2.1 PDF~\cite{NNPDF}
 sets
 and the comparison is shown in both the figures. The QCD
scales, both $\mu_R$ and $\mu_F$ are set to jet $p_T$ for inclusive
jet spectrum and average $p_T$ of two jets in case of di-jet spectrum.    
The NLO spectrum also corrected for non-perturbative effects which will be
discussed in Section 7.   

\section{Unfolding}
In any experimental measurement  one of the goal is to compare between data and theory predictions
or with results from other experiments. However in this measurement, finite detector resolution smear the 
physical quantities and as a consequence the measured observables are expected to differ 
from the corresponding true values. Therefore
in order to carry out a realistic comparison, it is required to make measured observable
free from any detector effects, namely unfold the data to be compared
with known theory prediction. Unfolding is a procedure used to get rid of all detector
level distortion from the measured spectrum. \newline
 
 It is a standard practice, for the measurement of a single observable,
 to apply $bin-by-bin$ correction in the measured observable. 
In this method one evaluates the generalized efficiency, which is defined 
as the ratio of number of events falling in a certain bin of the observed
spectrum to the number of events in the same bin in the true spectrum. 
The large number of simulated events are used to obtain this ratio for 
each bin of the observable. In general this efficiency can be larger 
than unity depending upon the number of events migrating away or towards 
a particular bin due to detector smearing. The major short falling of 
this method is that it fails to take care large bin migration of events and 
also it doesn't take into account the unavoidable bin to bin correlations
between adjacent bins. 
A method which overcomes this shortcomings is the regularized 
unfolding technique~\cite{Blobel:2002pu}. One of the most popular unfolding technique
of this class is Bayesian unfolding~\cite{UNF2}. The main essence of this 
algorithm lies with the treatment of different bins of the
true distribution as independent, i.e. without any correlation among each 
other and as a result it works for any kind of smearing. 
The core part of the algorithm is that it knows only about 'cause cells'
(number of events in a bin for true distribution) and 'effect cells'
(number of events in a bin for smeared distribution), but it doesn't 
know the position of the cells in the configuration space. The final 
goal of the problem is to estimate the probability of finding the 
true number of events in a given bin when the measured spectrum and 
some apriori knowledge on the detector smearing is available. 

\begin{figure}
\includegraphics[scale=0.35]{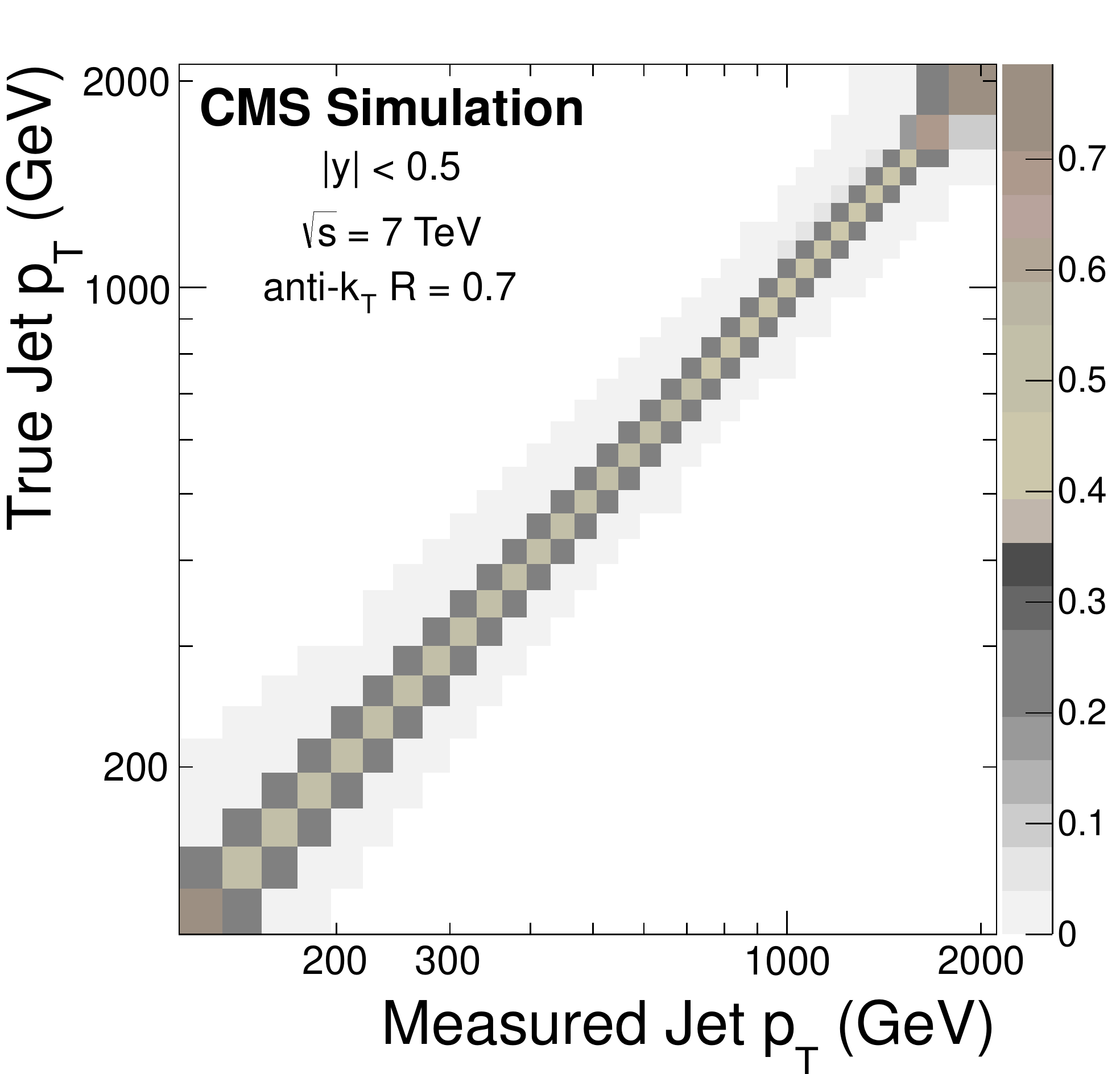}
\includegraphics[scale=0.35]{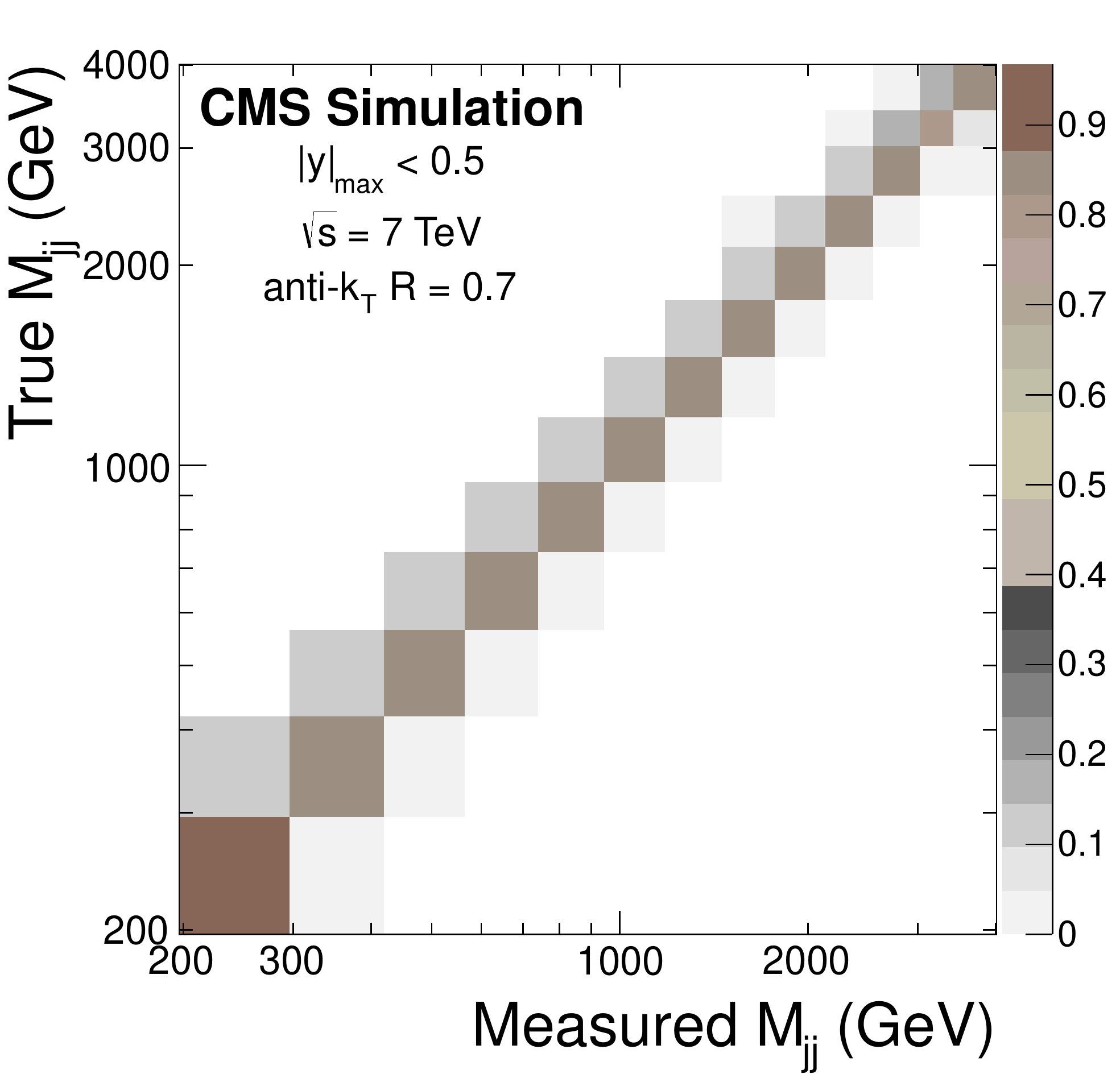}
\caption{Response matrix for the central rapidity bin for inclusive and dijet measurement.}
\label{fig:response_matrix}
\end{figure}

In the Bayesian unfolding process~(\cite{UNF2,UNF1}), the number 
of estimated 
events in the $i$-th bin of the unfolded distribution
('estimated causes') $\hat{n}(C_{i})$, as the result of applying the 
unfolding matrix $M_{ij}$ on the $j$-th bin of 
raw distribution ('effects'), containing $n(E_{j})$ events, is given by:
\br 
\hat{n}(C_{i}) = \sum_{j=1}^{n_{E}} M_{ij}n(E_{j}) ,
\er
where 
\br
M_{ij} = \frac{P(E_j|C_i)n_0(C_i)} {\epsilon_i \sum_{l=1}^{n_C}. 
P(E_j|C_l)n_0(C_l)}
\label{eq:mij}
\er 
Here $P(E_{j}|C_{i})$ is the $n_{E} \times n_{C}$ response matrix, 
where $n_{E}$ and $n_{C}$ are the number of bins in measured and unfolded 
distributions respectively. This response matrix causes correlation 
among different bins in the
unfolded distribution. Fig~\ref{fig:response_matrix} shows an example of 
response matrix for inclusive and dijet measurement used in the 7 TeV measurement.

Here $\epsilon_i$ = $\sum_{j=1}^{n_E} P(E_j|C_i)$ are 
efficiencies for each bin and $n_{0}(C_{l})$
is the number of entries in the $l$-th bin of the prior distribution. Output 
of each iteration of unfolding, goes as prior distribution to the next 
iteration.

Covariance matrix for unfolding is calculated by error propagation 
from $n(E_{j})$ which is denoted as $V(\hat{n}(C_{l}),\hat{n}(C_{m}))$. 
$M_{ij}$ is independent of $n(E_{j})$ for the first iteration only. So 
the error propagation from one iteration to another is described in 
form of a matrix which is given by~\cite{UNF1}, 
\br 
\frac{\partial \hat{n}(C_{i})}{\partial n(E_{j})} = M_{ij}+ 
\sum_{k=1}^{n_{E}}M_{ik}~n(E_{k})\bigg ( \frac{1}{n_{0}(C_{i})} 
\frac{\partial n_{0}(C_{i})}{\partial n(E_{j})}~-~ \sum_{l=1}^{n_{C}} 
\frac{\epsilon_{l}}{n_{0}(C_{l})}
\frac{\partial n_{0}(C_{l})}{\partial n(E_{j})} M_{lk} \bigg ). 
\er 

This depends upon the matrix elements 
$\frac{\partial n_{0}(C_{i})}{\partial n(E_{j})}$, which
is $\frac{\partial \hat{n}(C_{i})}{\partial n(E_{j})}$ from the previous 
iteration. For the first iteration, the L.H.S of the eq.~\ref{eq:mij} 
is purely $M_{ij}$. 
The covariance matrix of the unfolded distribution is obtained from 
the error propagation matrix as~\cite{UNF1}, 
\br 
V(\hat{n}(C_{k}), \hat{n}(C_{l})) = \sum_{i,j=1}^{n_{E}} 
\frac{\partial \hat{n}(C_{k})}{\partial n(E_{i})}~V(n(E_{i}),n(E_{j}))
\frac{\partial \hat{n}(C_{l})}{\partial n(E_{j})}.
\er 
from the covariance matrix of the measurement $V(n(E_{i}),n(E_{j}))$.

In the present study the goal is to compare $p_T$ and $M_{jj}$ spectrum
with the theory prediction which requires to unfold the corresponding
spectrum. Here in each $p_{T}$ or $M_{jj}$ bin is populated by jet 
yield that migrate from the neighboring bins. A fraction of yield from a
particular bin also migrate to the neighboring bins due to resolution
effects. In a steeply falling spectrum, like in the case of
QCD, the net effect is that for all the bins, number of jets migrate
in is much more than the number of events that migrate
out. 
In the process of unfolding, the first task is to construct a 
response matrix $\mathcal{R}$ of size $M \times N$, where $N$ is the 
number of bins in the measured spectrum, and the $p,q$ th element of 
this matrix gives the correlation between the $p$-th and $q$-th bin
of measured and true spectrum respectively. This response matrix 
$\mathcal{R}$ is operated on the measured spectrum to get the unfolded 
spectrum. The response matrix is created from simulation and the 
difference between data and simulation in terms of jet
energy resolution has been accounted. The unfolding is done by 
DiAgostini's iterative Bayesian unfolding method with iteration 
parameter equal to 4. Of course, the unfolded spectrum does not describe 
precisely the true spectrum, it represent it with certain uncertainty,
which will be discussed later.

 \section{Experimental Uncertainty}
The experimental uncertainty refers to all the uncertainties related with the measurements 
that affect the $p_T$ and $M_{jj}$ spectrum. The leading 
components of the
uncertainties are due to jet energy scale (JES), jet energy resolution (JER) and 
the luminosity uncertainty. The other uncertainty sources like
 angular resolution has very less effect on the measured spectrum. 
For the inclusive-jet measurement, the total relative uncertainty
from all the sources varies between $10\%$ to $35\%$ as we move from 
lower $p_{T}$ to higher while for the dijet mass measurement it varies between $5\%$ to $35\%$ for the entire
 mass range. On the other hand for both the measurements, the amount of relative uncertainty
 increases with the rapidity bins. 
The description of various uncertainty sources are given in
the following sub-section.

 \subsection{Jet Energy Scale Uncertainty}
The jet energy scale(JES) is the most dominant component of total uncertainty 
in the measurement involving $p_T$ of jets. Since we are dealing with a
very steeply falling spectrum in QCD($\sim ~ 1/p_{T}^{4}$), so a small 
uncertainty($\delta p_T$) in the measurement translates significant 
 uncertainty in the measurement of cross-section. The
JES uncertainty on the PFJets is of order of 
2\% to 2.5\% for the 7~TeV measurement.
The JES sources contain eleven mutually uncorrelated components, 
each representing a signed $1\sigma$ fluctuation from the central
value for each $p_{T}$ and $\eta$ bin and it is parametrized within the physically allowed 
$p_{T}$ and $\eta$ ranges~\cite{QCD-11-004}. Summing up the each 
contribution in quadrature gives the total
JES uncertainty.
The uncertainty sources are broadly divided into three categories. 
They are viz. PU effects, relative calibration of jet
energy scale  vs $\eta$ and absolute energy scale 
including $p_{T}$ dependence.

Although the JES uncertainty for the PU effects are mainly significant at 
very low $p_{T}$, but for 7 TeV measurement it is not so much significant.
The second broad category of uncertainty is due to L2 correction, mentioned in section 3.2.
 It parametrizes the possible variation of the JES,
and it is directly measured as the correlation between 
different $\eta$ values for a fixed $p_{T}$ bin. In general the uncertainties
can be $p_{T}$ dependent. However a detail study of data and MC showed that 
this dependence can be factorized without any loss of generality.  

 The last category is the absolute scale uncertainty or the uncertainty
introduced due to L3 correction, the most relevant one for the 
jet cross-section measurement analysis. The well measured $p_{T}$ balanced 
events are used to calibrate this $p_{T}$ dependent uncertainty, as mentioned in section 3.2. This calibration is 
used to constrain the JES in a finite $p_{T}$ region of $30-600~GeV$.
The JES beyond this regions are obtained from MC simulation. 
The  $p_{T}$ dependent uncertainty arising from modeling of
underlying event and jet fragmentation is obtained by comparing 
predictions from PYTHIA6~\cite{pythia} and HERWIG++~\cite{HERWIG}. The general studies show
that both the generators in general agree well with data. 
The uncertainty due to calorimetric response to hadrons is estimated
varying the response parametrization by $\pm3\%$ and 
comparing with the central value. 
The luminosity uncertainty for 7 TeV measurement is about $2.2\%$ 
which is $100\%$ correlated among different $p_{T}$ bins.

\begin{figure}
\includegraphics[scale=0.35]{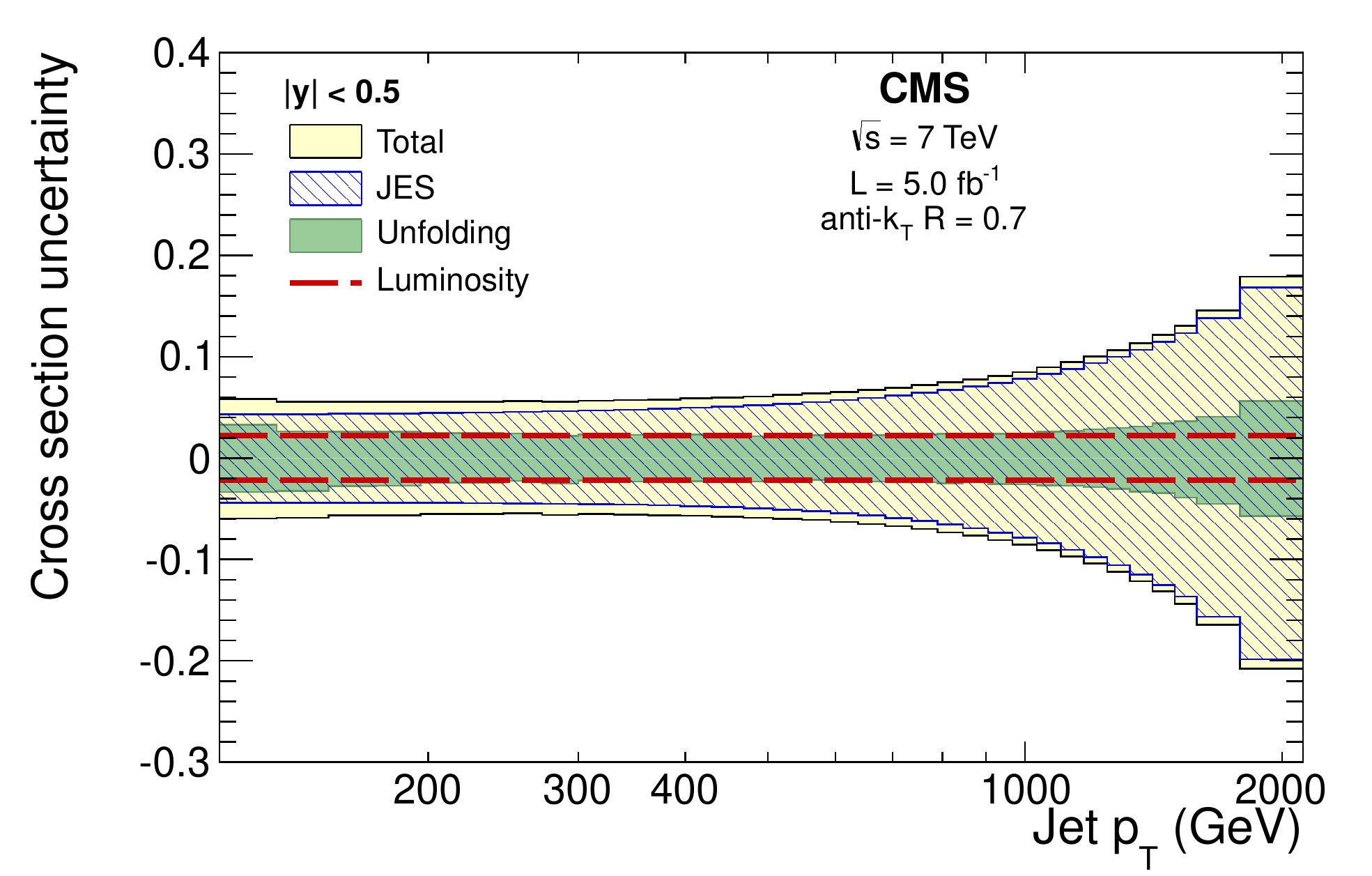}
\includegraphics[scale=0.35]{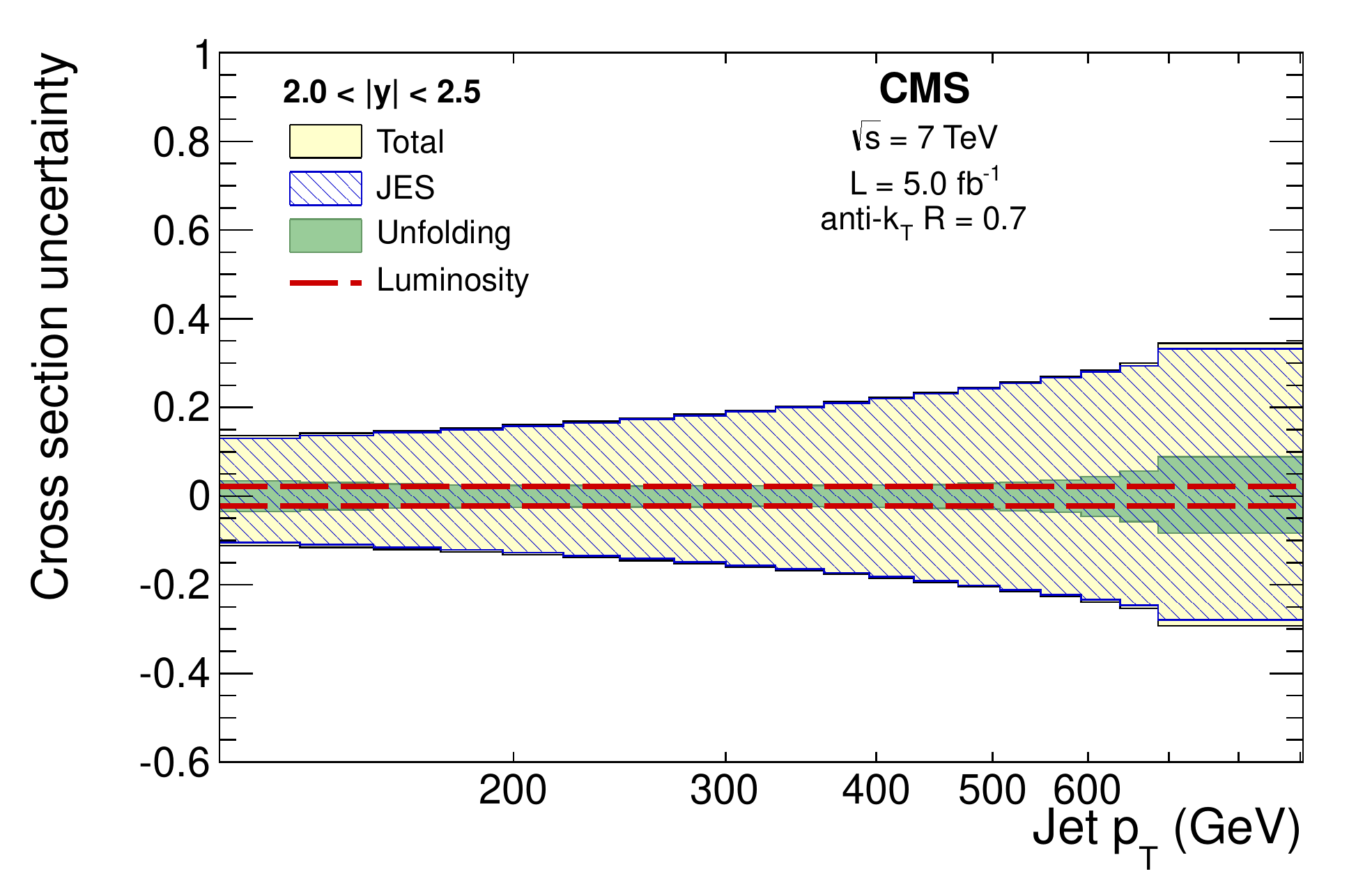}\\
\includegraphics[scale=0.35]{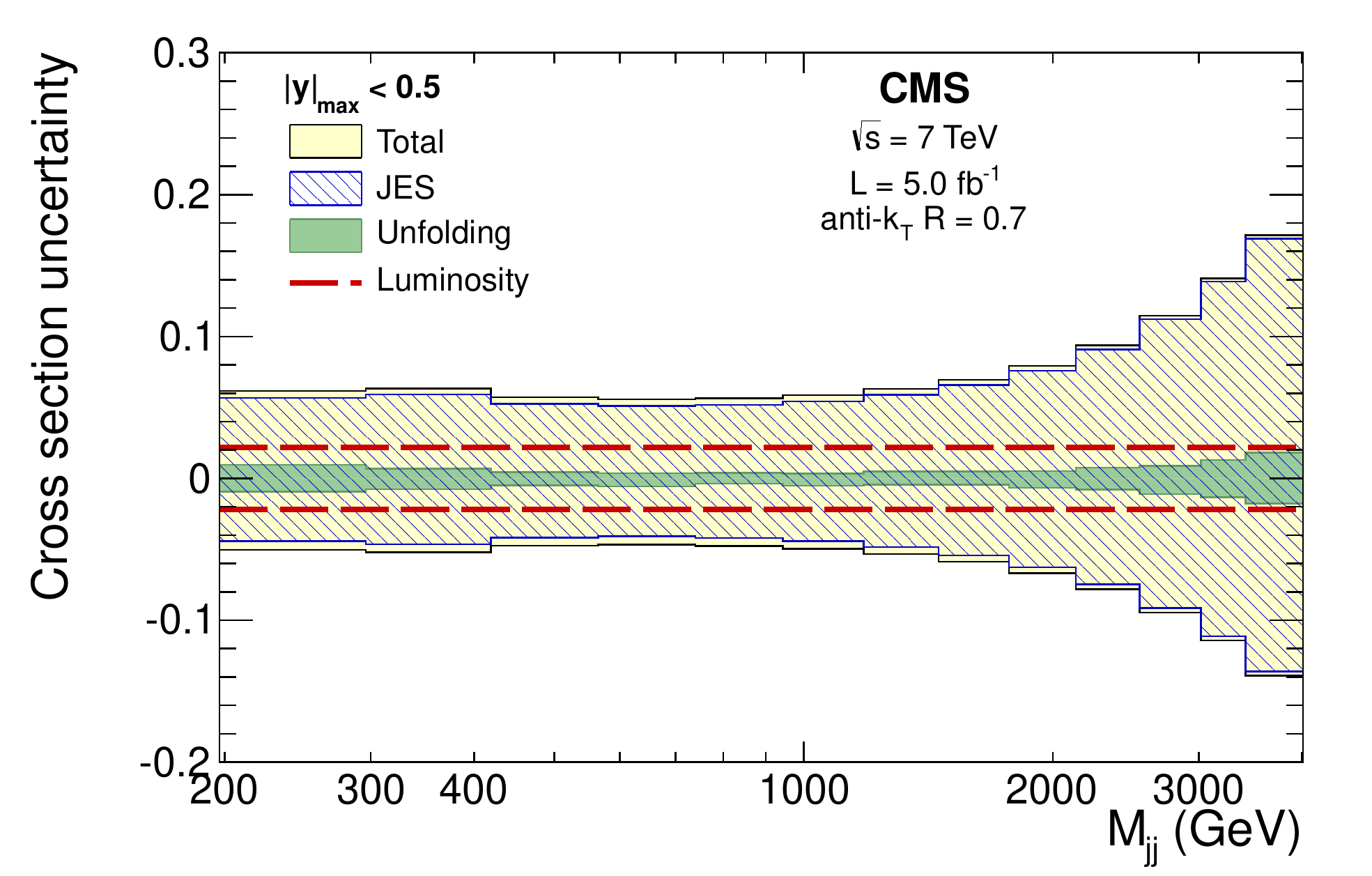}
\includegraphics[scale=0.35]{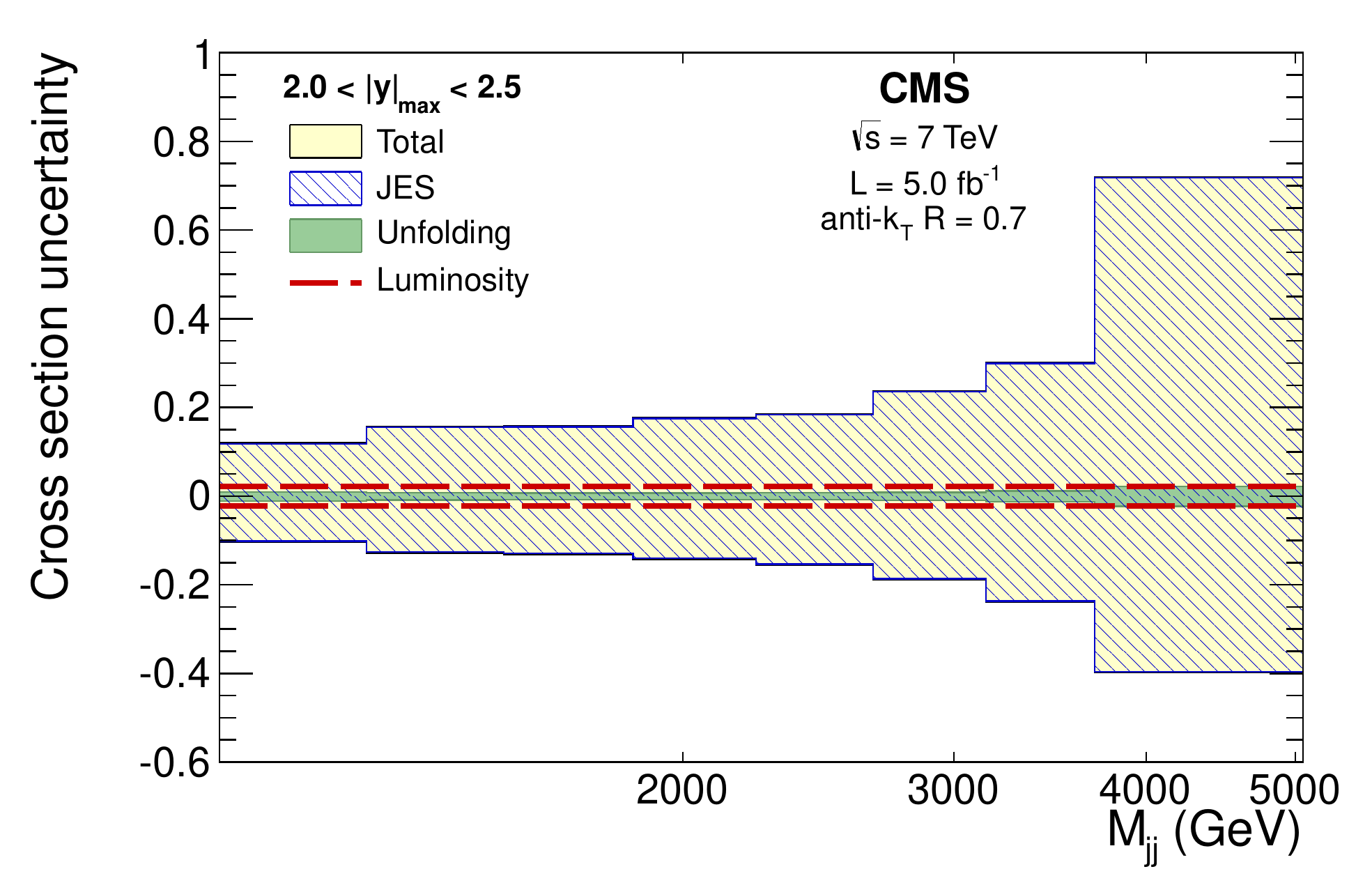}
\caption{Relative experimental uncertainties for inclusive jet and Dijet mass measurement.}
\label{fig:exp_unc}
\end{figure}

 Fig~\ref{fig:exp_unc} shows the uncertainties in the cross section measurement 
due to different components of total systematics for a wide range of $p_{T}$ and $M_{jj}$. As 
mentioned before, the most dominant contribution to the total uncertainty in the cross section is due to JES.
In the measurement of inclusive jet cross section, the total uncertainty is about 5-20\% (10-30\%), where as
for dijet mass measurement it varies between 5-30\% (10-60\%) for central (outermost) rapidity bin in both the cases.

 \subsection{Unfolding Uncertainty}
 The unfolding correction is directly related to $p_{T}$ and $M_{jj}$ 
resolution. In case of dijet mass the resolution varies between
$2\%$ to $5\%$ in all rapidity bins and becomes larger as the 
dijet mass increases.  For inclusive jet, the same
varies between $5\%$ to $10\%$ and the uncertainty increases with 
increasing jet $p_{T}$.

 \section{Theory Calculation}
Once the experimental measurement is performed, the next task is 
to compare the theory predictions with the unfolded measured spectrum.
 It is reasonable and relatively precise to use NLO
calculation to predict theoretical values of cross section.
Furthermore, NP correction factor is applied to account
for effect due to hadronization and multiparton interaction
 (MPI). These are discussed briefly in the next section.

 \subsection{NLO Calculations}
 In hadron collision, the QCD jet cross sections are 
available for a long time ~\cite{Ellis:1992en,Kunszt:1992tn,Giele:1993dj,Trocsanyi:1996by}. 
The recent NLO calculation for three jet observables
are also performed~\cite{Nagy:2003tz}.
Here the NLO jet cross-sections are estimated using
NLOJet++ which is based on the calculation described in Ref.~\cite{Nagy:2003tz}.
However, since the NLO calculation is very much time consuming, it is
optimized within the framework of fastNLO(v1.4)~\cite{Kluge:2006xs}
package. In calculating jet cross sections one needs to provide PDF
and QCD scales which are not exactly defined. As a consequence,
theory calculations are subject to some uncertainties due to the various choices of
PDFs and QCD scales.
The renormalization ($\mu_{R}$) and factorization ($\mu_{F}$) scales
for the inclusive and dijet measurements, are set equal to the
jet $p_{T}$ and the average transverse momentum  of the
two jets($p_{T}^{Avg}=\frac{p_{T}^{1}+p_{T}^{2}}{2}$), respectively.
The NLO calculation is done with five different
PDF sets: CT10~\cite{CT10}, MSTW2008NLO~\cite{MSTW}, NNPDF2.1~\cite{NNPDF},
HERAPDF1.5~\cite{HERA}, and ABKM09~\cite{ABKM} at the
corresponding default values of the strong coupling
constant $\alpha_{S}(M_{z})$= 0.1180, 0.120, 0.119, 0.1176, and 0.1179,
respectively.

 \subsection{Non Perturbative Correction}
As explained before in p-p collision, due to hard interaction colored
partons(quarks and gluons) are produced and subsequently they undergo cascades and produce
various colored combination of quarks and anti-quarks
which finally recombine among themselves to produce colorless hadrons
which hit the detector. This is a process called fragmentation or
hadronization. These colorless hadrons, primarily K and $\pi$ mesons enter into 
detector and deposit energy in its different components. In contrast to hard interaction which is a short distance effect,
this process is a long distance effects and occurs at very low energy
scale where the strong couplings constant $\alpha_s$ becomes too large
and perturbation theory fails. Hence, the hadronization is pure
non-perturbative effects which cannot be estimated using techniques
of pQCD. Moreover, in p-p collision the hard partons from protons which
share a fraction of initial beam energy produce hard scattered events 
leaving the rest of the partons as beam remnant of protons. Therefore,
any hard interaction is also accompanied by many soft interactions among
this remnant of protons, which is known as multiple parton interaction (MPI).
Note that MPI also cannot be described from the first principles of QCD, it is also a purely
NP effect. Therefore, parton level cross sections are required to be
corrected to obtain particle level cross section.
This non perturbative(NP) correction factor accounts for the amount of
correction required due to hadronization and MPI bin-by-bin.
This correction is estimated by taking the ratio of the cross section
predicted with the nominal settings for MPI and hadronization
model parameters and the cross section obtained without any effects
of MPI and hadronization. The NP correction factor is defined as 
$c_{NP}~ =~ \frac{\sigma(Nominal)}{\sigma(NoMPI, NoHAD)}$. The
numerator is the cross-section with the nominal settings for MPI and hadronization in the generators, 
where as the denominator represents the same without any of these effects. \\
These are estimated from MC simulation using
two different generators\\
 PYTHIA6(tune Z2)~\cite{pythia} and
HERWIG++~\cite{HERWIG}. The chosen MC models are representative of the
possible values of the non-perturbative corrections, due to their
different physics description. The average value of the NP correction
factor is estimated as the mean value from two different generator for
each $p_{T}$ and $M_{jj}$ bin. The uncertainty due to NP correction is
estimated as the difference between the values obtained from two
different generators, for each bin. Eventually to obtain
the full theory spectrum correctly, NP correction factor
is multiplied with the NLO theory prediction. In this measurement, the NP
correction varies from 1\% to 20\% ~\cite{QCD-11-004}.

\subsection{Theoretical Uncertainty}
The variation among different sets of PDF introduces a total
uncertainty up to $30\%$ on the theoretical prediction of the
cross-section for both inclusive-jet and dijet measurements.
On the other hand,  the variation of $\alpha_{S}(M_{z})$ by 0.001
translates $1-2\%$ uncertainty on total cross-section.
The scale uncertainty is determined by varying the renormalization and
factorization scale for six different points
($\mu_F/\mu,\mu_R/\mu) =(0.5,0.5)$, $(2,2)$, $(1,0.5)$, $(1,2)$, $(0.5,1)$,
$(2,1)$, where $\mu= Jet~p_{T}$ for inclusive jets
and $\mu =p_{T}^{Avg}$ for dijet measurement.

\begin{figure}
\includegraphics[scale=0.40]{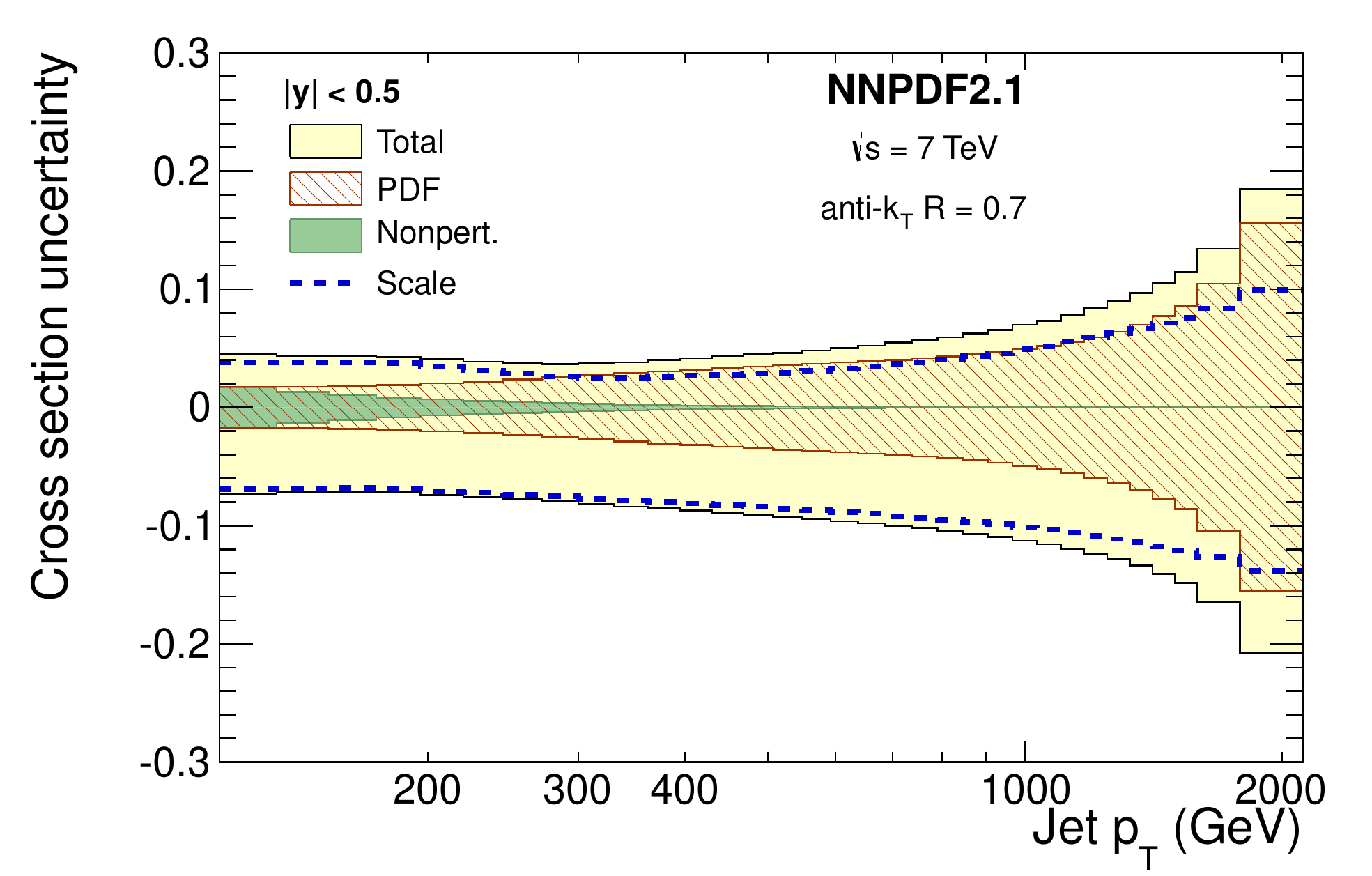}
\includegraphics[scale=0.40]{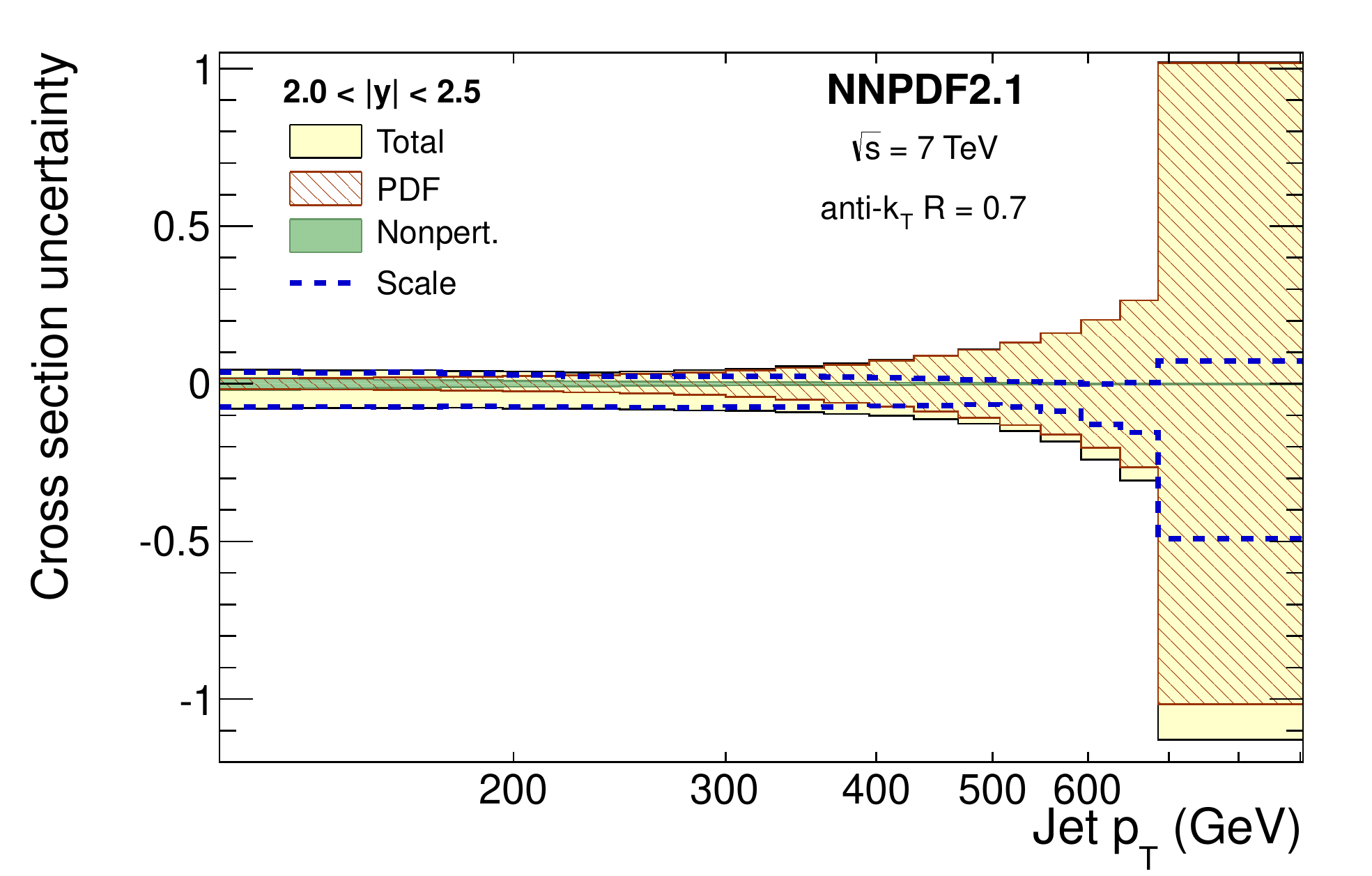}\\
\caption{Different components of total theoretical uncertainty for inclusive jet,
for two different rapidity bins.}
\label{fig:incl}
\end{figure}

\begin{figure}
\includegraphics[scale=0.40]{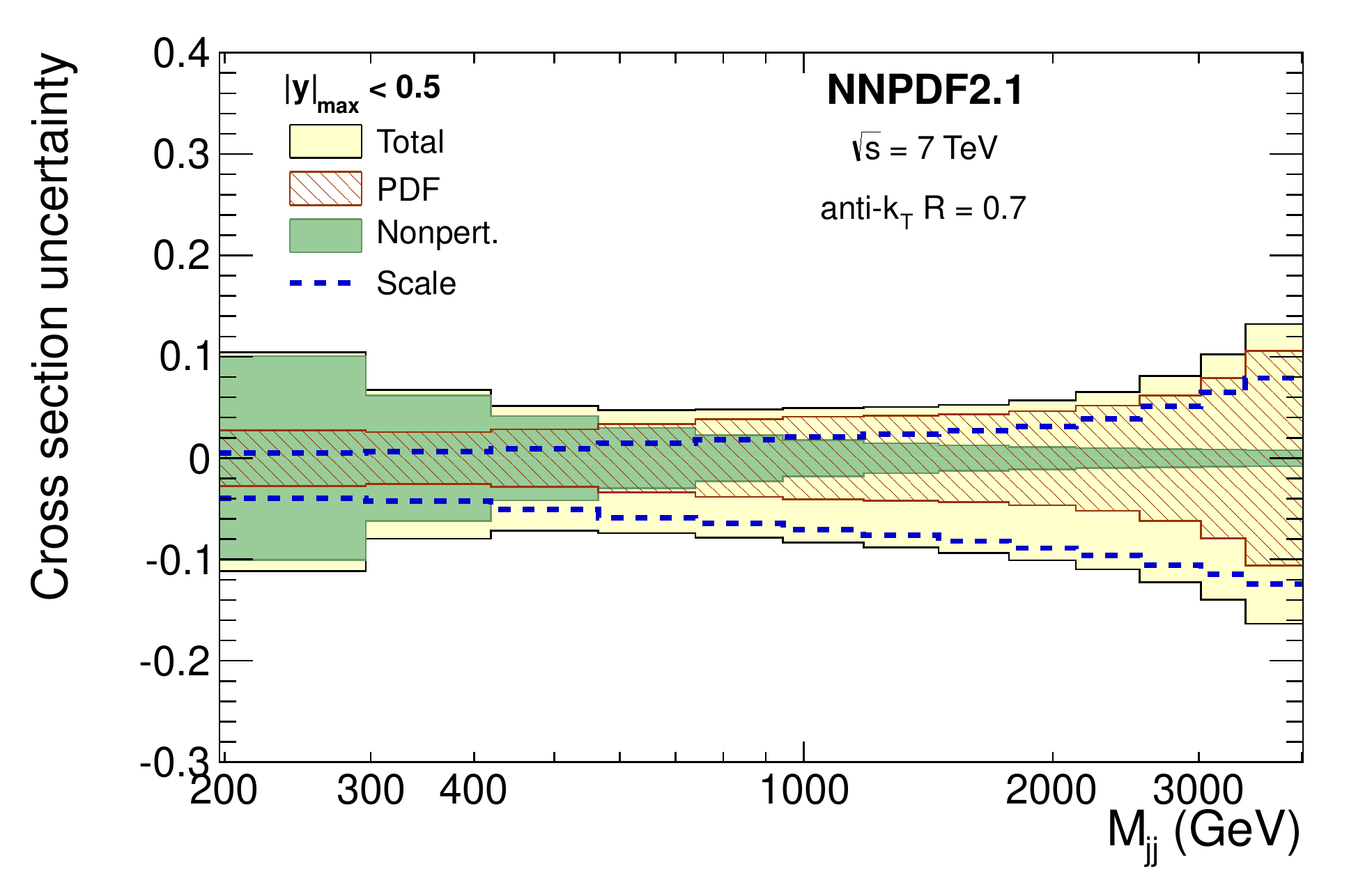}
\includegraphics[scale=0.40]{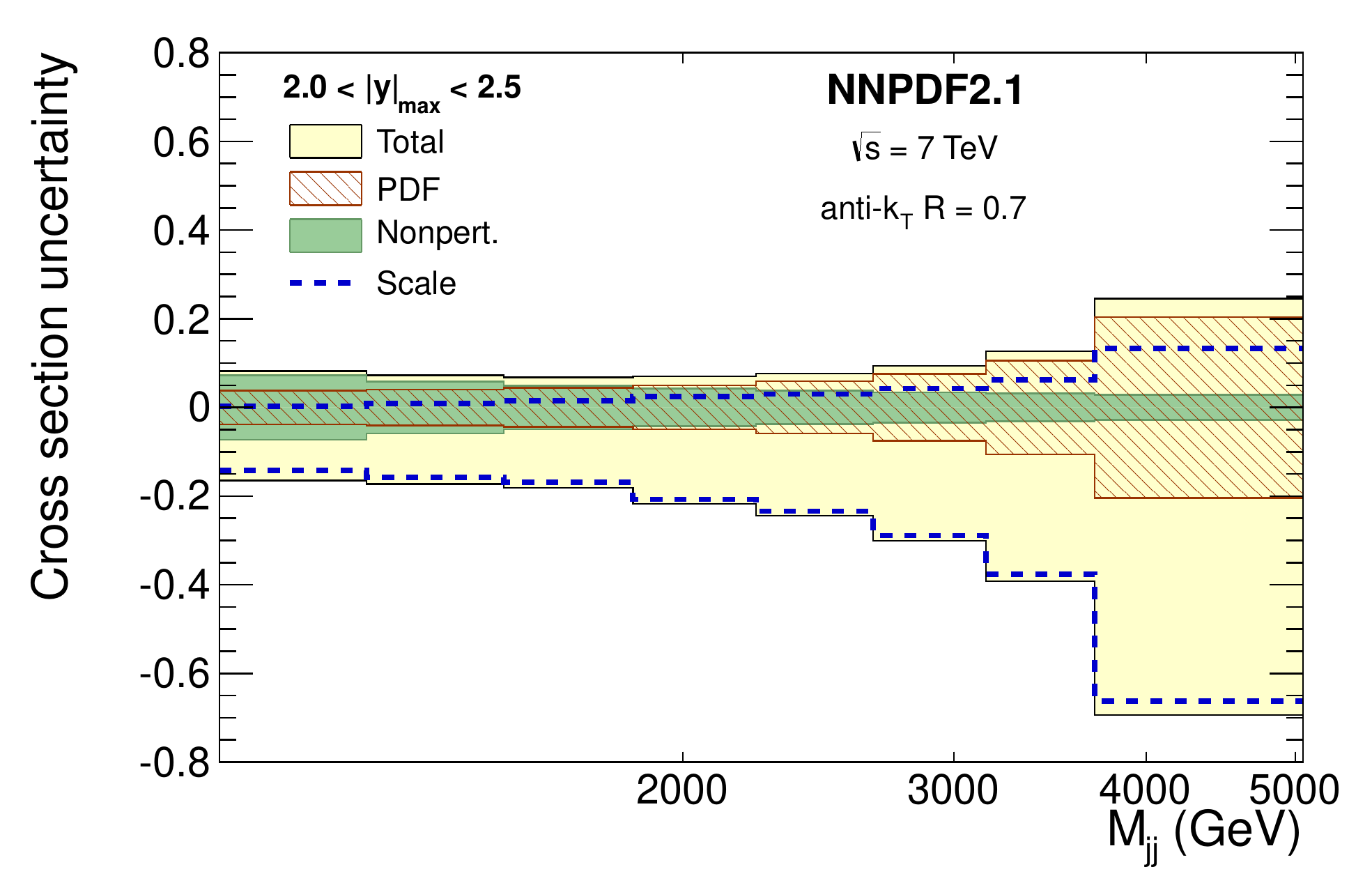}\\
\caption{Different components of total theoretical uncertainty for dijet jet,
for two different rapidity bins.}
\label{fig:dijet}
\end{figure}

In Fig.~\ref{fig:incl} and Fig.~\ref{fig:dijet}, we demonstrate the total uncertainty on the
theory cross section calculation due to different sources namely PDF, QCD scale and NP for a wide
range of $p_T$ and for two extreme bins of rapidity, taking NNPDF2.1 as the 
benchmark PDF. Note that for the barrel region,
the uncertainty is within 5\% for the range of $p_T$ upto $\sim$ 1 TeV and
thereafter it increases. In case of outer most rapidity bin, uncertainty
becomes large(more than 10\% or above) for $p_T$ beyond 500~GeV.
Notice that, uncertainty due to PDF dominates over the
others. Similarly, in Fig.~\ref{fig:dijet} the uncertainties on dijet cross section predictions are shown
for various $M_{jj}$ values for two rapidity bins. In this case the uncertainty due to NP corrections large at
lower bins of $M_{jj}$. Evidently, PDF uncertainty contributes most to the total theoretical uncertainty.

\section{Results}
The inclusive jet cross section in terms of double differential measurement
are compared with data for five different PDF sets as mentioned before
and for five rapidity bins~\cite{QCD-11-004}.
\begin{figure}
\includegraphics[scale=0.40]
{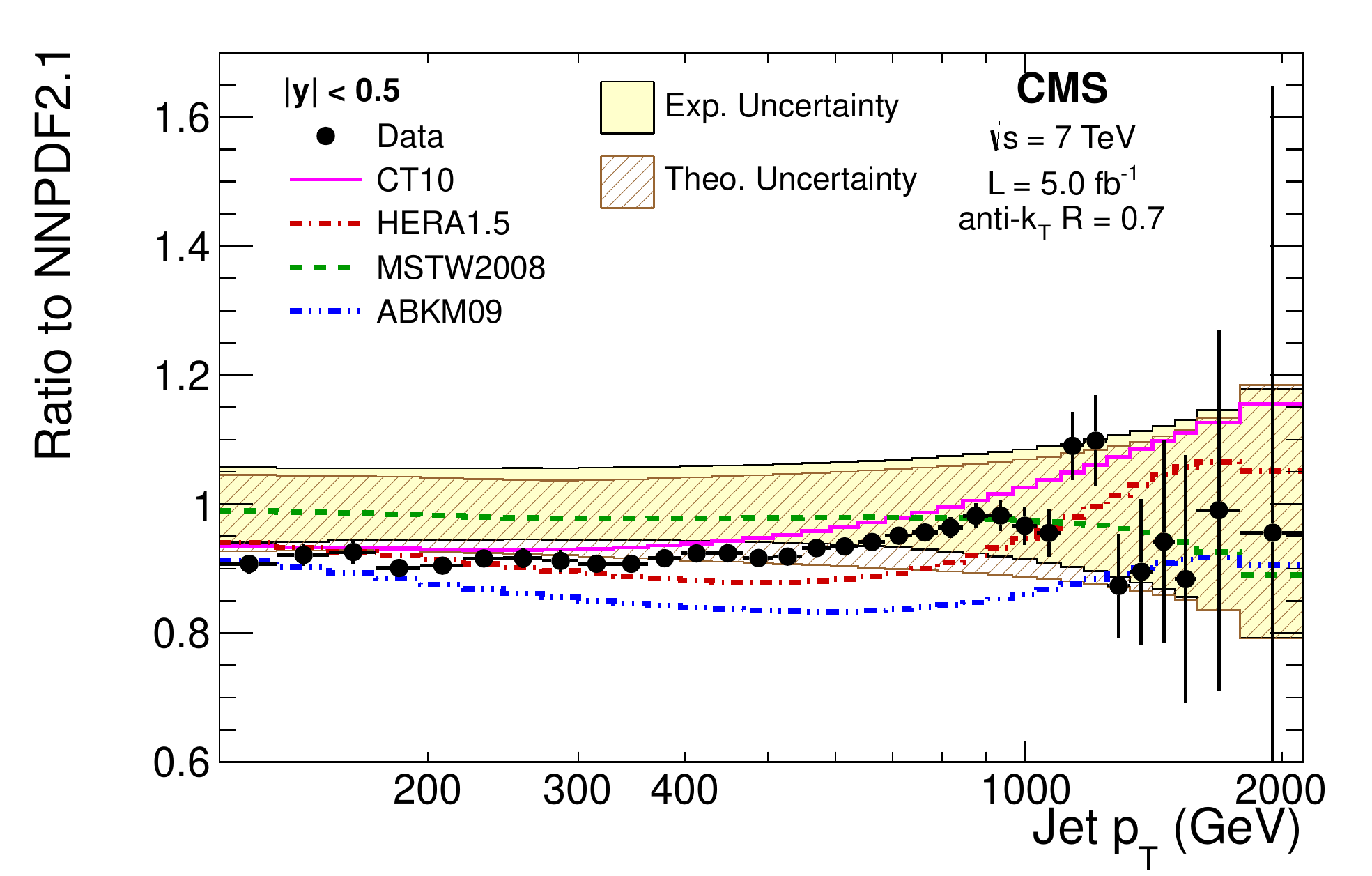}
\includegraphics[scale=0.40]
{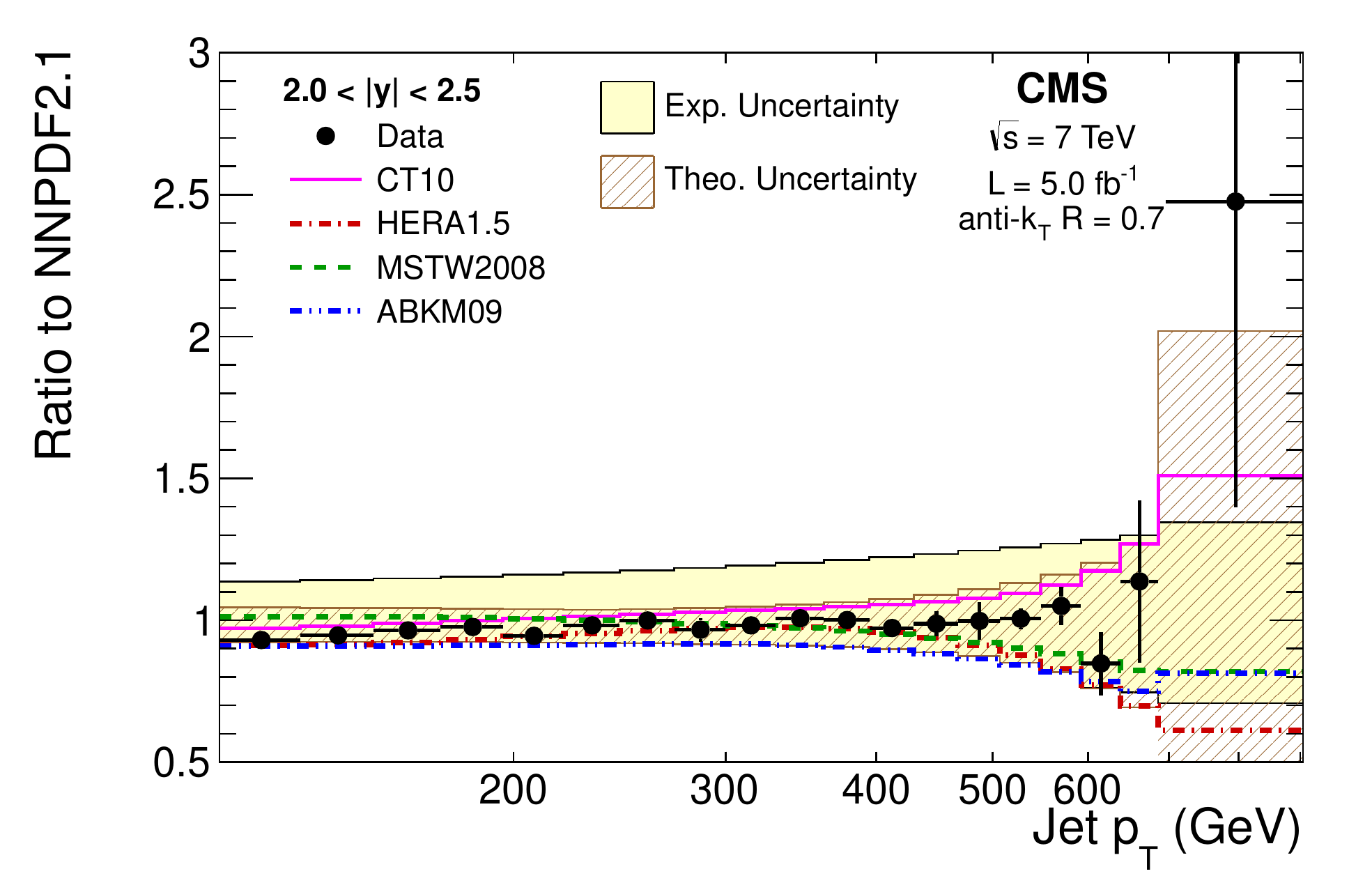}\\
\caption{Data over theory for NNPDF2.1 theory
prediction of inclusive jet $p_{T}$.
for two extreme rapidity bins.}
\label{fig:DT_incl}
\end{figure}

\begin{figure}
\includegraphics[scale=0.40]
{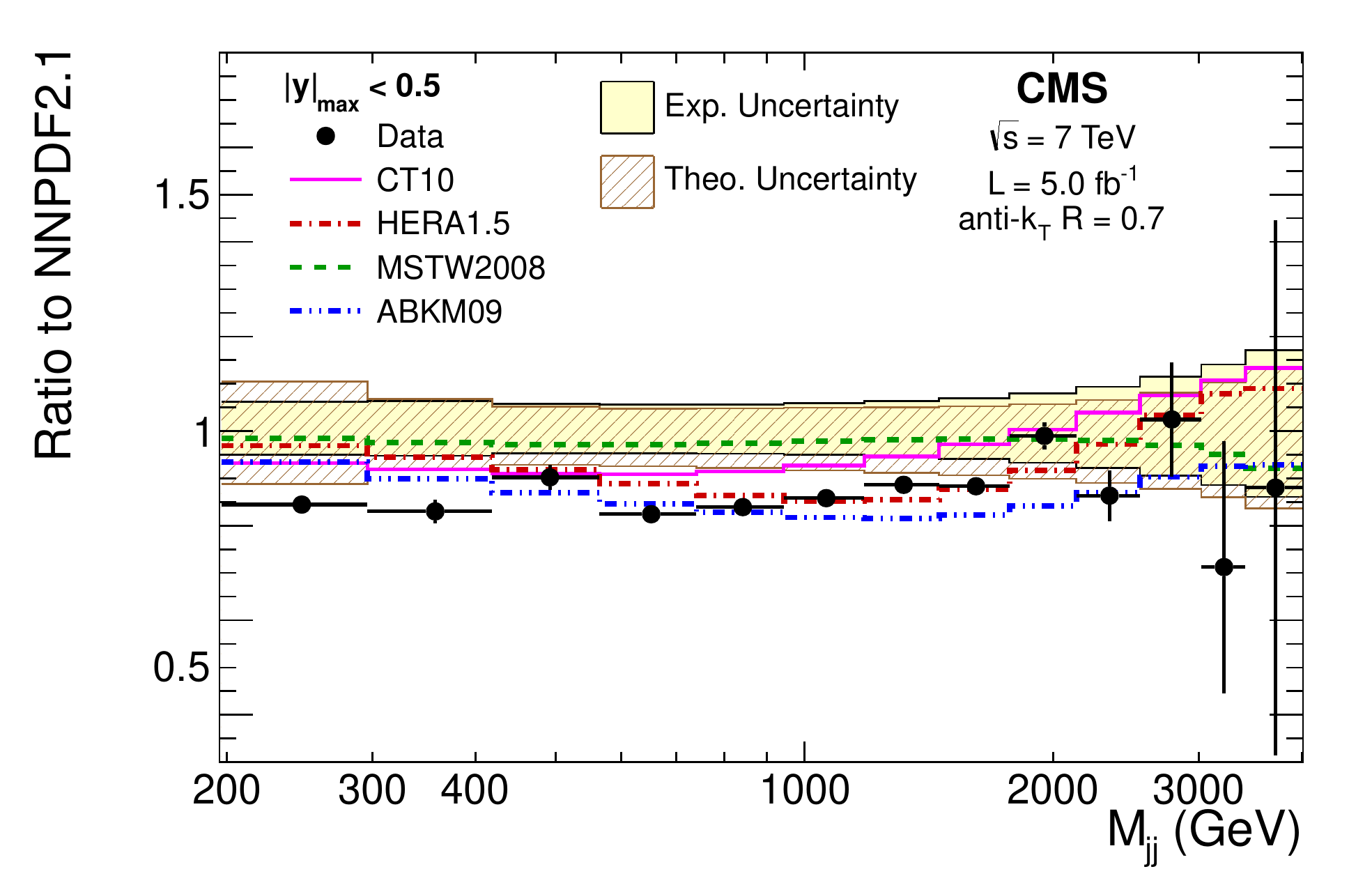}
\includegraphics[scale=0.40]
{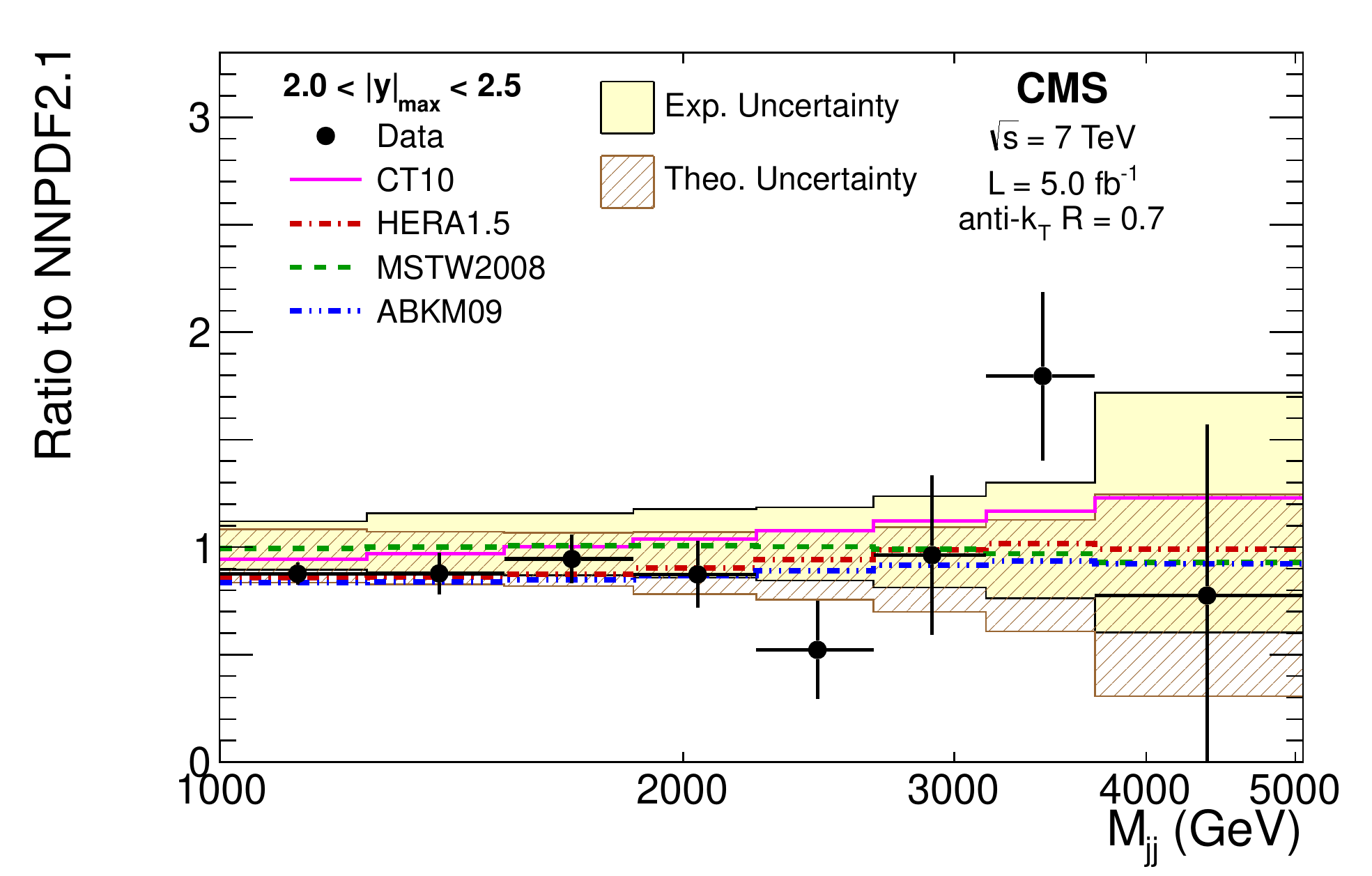}\\
\caption{Data over theory for NNPDF2.1 theory
prediction of dijet mass for two extreme rapidity bins.}
\label{fig:DT_dijet}
\end{figure}
For the sake of illustration, in Fig.~\ref{fig:DT_incl} and
~\ref{fig:DT_dijet} the theory prediction based on
 NNPDF2.1 is compared with data for both inclusive and dijet cases and are shown
for two extreme rapidity bins, viz. $0.0<|y|<0.5$ and $2.0<|y|<2.5$,
 along with total experimental and theoretical uncertainties.
Note that the total experimental uncertainty is comparable with the
theoretical uncertainty.
Clearly, in both cases, this ratio is within
the band of uncertainty($\sim 8-10$\%). The error
bars are too large for high $p_T$ or $M_{jj}$ range, and is dominantly due
to statistical uncertainty. Similar pattern of agreement is observed for other PDF sets
as well, except for ABKM09 set, in which case this ratio goes out of
the error band for the low rapidity bins, $|y|<$1~\cite{CMS-PAS-QCD-11-004}. However, a good
agreement between theory and data is observed as presented in Fig.~\ref{fig:DT_incl},\ref{fig:DT_dijet}.

The detail studies are also carried out to understand data and theory
compatibility by comparing their ratio using the central
value of each PDF sets. In Fig.~\ref{fig:DT_incl} and
Fig.~\ref{fig:DT_dijet}, we demonstrate this comparison for
NNPDF2.1 PDF set for inclusive and dijet mass measurement respectively.
In the same Figures the other curves represent the
ratio between the calculations based on NNPDF2.1 and other
PDF sets, like ABKM09, CT10, HERAPDF1.5, MSTW2008NLO as shown.
Here the results are presented for two rapidity bins, however, the studies
are done for all rapidity bins~\cite{CMS-PAS-QCD-11-004}. In case of inclusive
jet data, for central region, the agreement for NNPDF2.1 PDF set is within 5-10\%
for a wide range of $p_T$, but for higher $p_T$($>$1 TeV),
it is affected by large uncertainty. This conclusion remains true also for 2.0$<|y|<$2.5 bin,
where the total uncertainty is too large, in particular for high $p_T$
range. Similar level of agreement is also observed for di-jet case and
but, at larger values of $M_{jj}$, the uncertainty is too large, so it is far from
any conclusion to be drawn. In both cases more or less similar type of
behavior of ratio measurements are observed.

\section{Summary}
The inclusive jet cross section and dijet invariant mass measurement
are reviewed here with some detail discussions. In this measurement the
important quantities to be measured are the momentum and energy of 
jets which is a very non trivial objects from both theoretical and
experimental perspective. Before describing this measurement we
discuss various issues related with jets. In the formation of jets,
the main constituents are either four momenta of reconstructed stable particles
or four momenta of calorimeter towers which are combined to obtain the 
kinematic properties of jets. The combination methods of momentum of jet 
constituents are guided by certain theoretical prescriptions, which take into account
singularities, present due to the very soft or collinear branching of the partons. 
In order to 
avoid these theoretical problems, the jet algorithms are properly designed,
resulting the evolution of various types of jet algorithms. 
In this context the anti-$k_T$ jet algorithm is also discussed, which is currently the most widely
used algorithm in hadron collider experiments.  \newline

The different techniques adopted in CMS for jet reconstruction and JES are 
discussed. 
Due to the non linear response of the detector, the measured jet energy 
requires to be corrected. The various sources of this correction factors
are presented and it is found that the derived JEC
from $\sqrt{s}=7$ TeV and $\int L$ = 5${\rm fb}^{-1}$ data is at-most 20\%.
In performing this measurement the details of the selection of good quality
of events and jets are presented including a short description
of implemented trigger system in CMS.
Finally the results of this measurement are presented in terms of double
differential cross section for a wide range of $p_T$ and $\rm {M}_{jj}$ for a five regions
of rapidity bins covering $0 \le{\rm |y|} \le 2.5$ with an interval $\Delta |y|~ =$ 0.5. \newline
 
Realistically, to make a comparison of this measurements with theory 
prediction or results obtained
by other experiments, it is required to eliminate smearing effects 
of the detector in
the measurement, which is a delicate task. Thus, the unfolding
of data, the corresponding algorithm and how it is implemented in this particular measurement
are described in details. Moreover, measurements are also affected by different type of
systematics including statistical uncertainties, which all are need
to be accounted. It has been observed that the main source of uncertainty is 
primarily due to JES, which varies between 2-2.5\% and translates an
uncertainty 5-35\% on the measured inclusive jet cross section across varying
$|y|$ and $p_{T}$ bin. The
same JES uncertainty for the dijet mass cross section shoots up to 60\% for the outermost rapidity bin.\newline

 On the other hand, the theory
prediction based on LO calculation is not adequate to describe 
this process. Hence the theoretical estimations are obtained based on NLO 
calculation and is performed for five different PDF sets viz. 
ABKM09, HERA15, CT10, MSTW2008 and NNPDF2.1. Of course, the various sources
of uncertainties in these predictions are mentioned and it is observed
that choice of various PDF is the dominant source of theoretical uncertainty
which is about 30\%  on calculated cross section. 
 In addition, a non-perturbative
correction, arising due to the multiple parton interaction and hadronization effects, 
is applied to these NLO theory prediction.
Eventually, the comparison between data and theory predictions are presented 
 along with the total theoretical and experimental uncertainty limits.
It is found both the measurements, inclusive jet and dijet mass cross sections, and theory predictions agree within
10\% and this deviation is in general covered by the total theoretical and experimental
uncertainty limits. \newline

 This measurement confirms once more the success of pQCD with a 
very high precision unambiguously 
at the probed energy regime. This measurement probes a wide region in phase-space.
The inclusive jet $p_{T}$ varies between 100 GeV to 2 TeV, whereas the dijet invariant mass extends up to
5 TeV. The momentum fraction carried by the partons, $x$, probed in this experiment  $0.03<x<0.57$. 
The application of this measurements are very wide, for instance, 
it can be used for extraction of the strong coupling constant 
$\alpha_{S}$ and testing its evolution with energies as predicted by the theory
QCD. Furthermore, this measurement can be used to 
constrain the parameters of different PDF sets as well. \newline

{\large \bf{Acknowledgements}}\\
The authors acknowledge the CMS collaboration and the India-CMS collaboration for their support. They
are also grateful to  Bora Isildak, Mithat Kaya, Ozlem Kaya, Konstantinos Kousouris, Klaus Rabbertz, Mikko Voutialinen,
 Niki Saoulidou, Maxime Gouzevitch, Sung-Won Lee and Jeffrey Berryhill
for discussion on several occasion. 

\bibliographystyle{utphys}
\bibliography{Jets-v4}{}
\end{document}